\documentclass[lettersize,journal]{IEEEtran}
\usepackage{amssymb}
\usepackage{cite}
\usepackage{graphicx}
\usepackage{array}
\usepackage{multirow}
\usepackage{amsmath}
\usepackage{stfloats}
\usepackage{graphicx}
\usepackage{subfigure}
\usepackage{tabularx}
\usepackage{epsfig,epsf,balance,cite}
\usepackage[caption=false,font=normalsize,labelfont=sf,textfont=sf]{subfig}
\usepackage{setspace}
\usepackage{bm}
\usepackage{textcomp}
\usepackage{algorithmic}
\usepackage{algorithm}
\usepackage{subfigure}
\usepackage{caption}
\usepackage{graphicx}
\usepackage{setspace}
\usepackage{url}
\usepackage{verbatim}
\usepackage{graphicx}
\usepackage{cite}
\usepackage{amssymb}
\usepackage{color}
\usepackage{xpatch}
\definecolor{shadecolor}{RGB}{0,0,255}
\definecolor{blue}{RGB}{0,0,255}

\newtheorem{theorem}{Theorem}

\newtheorem{lemma}{Lemma}

\makeatletter

\ExplSyntaxOn
\cs_new:Npn \bibColoredItems #1#2
{
	\clist_map_inline:nn {#2} { \cs_new:cpn {bib@colored@##1} {#1} } 
}
\ExplSyntaxOff

\newcommand\bib@setcolor[1]{%
	\ifcsname bib@colored@#1\endcsname
	\expanded{\noexpand\color{\csname bib@colored@#1\endcsname}}%
	\else
	\normalcolor
	\fi
}

\xpatchcmd\@bibitem
{\item}
{\bib@setcolor{#1}\item}
{}{\fail}

\xpatchcmd\@lbibitem
{\item}
{\bib@setcolor{#2}\item}
{}{\fail}
\makeatother

\begin{document}

\title{Resource Allocation for Cell-Free Massive MIMO-aided URLLC Systems Relying on Pilot Sharing}

\author{Qihao Peng, Hong Ren,~\IEEEmembership{Member,~IEEE}, Mianxiong Dong,~\IEEEmembership{Member,~IEEE},\\ Maged Elkashlan, \IEEEmembership{Senior Member,~IEEE}, Kai-Kit Wong,~\IEEEmembership{Fellow,~IEEE},\\ and Lajos Hanzo,~\IEEEmembership{Life Fellow,~IEEE}
\thanks{Manuscript received September 12, 2022; revised January 2, 2022; accepted March 7, 2023.}
\thanks{Q. Peng and M. Elkashlan are with the School of Electronic Engineering and Computer Science at Queen Mary University of London, U.K. (e-mail: \{q.peng, maged.elkashlan\}@qmul.ac.uk). H. Ren is with National Mobile Communications Research Laboratory, Southeast University, Nanjing, China. (e-mail: hren@seu.edu.cn). M. Dong is with the Department of Information and Electronic Engineering, Muroran Institute of Technology, Muroran 050-8585, Japan (e-mail: mxdong@mmm.muroran-it.ac.jp). K. K. Wong is with the Department of Electronic and Electrical Engineering, University College London, London WC1E 7JE, U.K. (email: kai-kit.wong@ucl.ac.uk). L. Hanzo is with the School of Electronics and Computer Science, University of Southampton, SO17 1BJ, Southampton, UK. (email: lh@ecs.soton.ac.uk).(\emph{Corresponding author: Hong Ren}.) }
\thanks{The work of H. Ren was supported in part by the National Natural Science Foundation of China (Grant No. 62101128), and in part by Basic Research Project of Jiangsu Provincial Department of Science and Technology (Grant No. BK20210205). L. Hanzo would like to acknowledge the financial support of the Engineering and Physical Sciences Research Council projects EP/W016605/1 and EP/X01228X/1 as well as of the European Research Council's Advanced Fellow Grant QuantCom (Grant No. 789028). The work of Q. Peng was supported by the China Scholarship Council. }
}

\maketitle

\vspace{-0.4cm}
\begin{abstract}

Resource allocation is conceived for cell-free (CF) massive multi-input multi-output (MIMO)-aided ultra-reliable and low latency communication (URLLC) systems. Specifically, to support multiple devices with limited pilot overhead, pilot reuse among the users is considered, where we formulate a joint pilot length and pilot allocation strategy for maximizing the number of devices admitted. Then, the pilot power and transmit power are jointly optimized while simultaneously satisfying the devices' decoding error probability, latency, and data rate requirements. Firstly, we derive the lower bounds (LBs) of ergodic data rate under finite channel blocklength (FCBL). Then, we propose a novel pilot assignment algorithm for maximizing the number of devices admitted. Based on the pilot allocation pattern advocated, the weighted sum rate (WSR) is maximized by jointly optimizing the pilot power and payload power. To tackle the resultant NP-hard problem, the original optimization problem is first simplified by sophisticated mathematical transformations, and then approximations are found for transforming the original problems into a series of subproblems in geometric programming (GP) forms that can be readily solved. Simulation results demonstrate that the proposed pilot allocation strategy is capable of significantly increasing the number of admitted devices and the proposed power allocation achieves substantial WSR performance gain.
\end{abstract}


\section{Introduction}
Ultra-reliable and low-latency communication (URLLC) is one of the critical techniques for enabling the wireless control of the Industrial Internet-of-Things (IIoT) devices in smart factories \cite{2014Industrial,2019URLLC}, \textcolor{black}{where packets conveying on the order of a few hundred bits are delivered \cite{ref5}}. According to Shannon's capacity theory, the decoding error probability (DEP) approaches zero, when the blocklength tends to infinity \cite{1948Shanon}. However, in URLLC, the channel blocklength is limited, where the DEP typically fails to approach zero, which should be taken into account in the system design. The expression of the achievable data rate of short-packet transmissions was derived in \cite{ref3,2017nonconvex}, which analyzed the impact of finite channel blocklength (FCBL), signal-to-noise ratio, and the DEP on the data rate. Additionally, in smart factories, the wireless systems should simultaneously support a large number of devices such as robots and actuators with high reliability in the short channel blocklength regime, which is a challenging task.

Recently, massive multiple-input and multiple-output (mMIMO) enabled URLLC has attracted extensive attention, since it can simultaneously support multiple devices without sacrificing the time-frequency resources \cite{ref11,2014MIMO,ref13a,ref13d,ref13,zhao2021achieving}. It has been shown that mMIMO systems can support multiple devices even in the face of severe shadow fading channels \cite{ref13a}. The weighted sum rate (WSR) was maximized by jointly optimizing the pilot power and payload power using idealized channel state information (CSI) in \cite{ref13d}. The impact of realistic imperfect CSI and pilot contamination on the data rate were analyzed in \cite{ref13}. Then, the energy efficiency of mMIMO-aided URLLC was studied in \cite{zhao2021achieving}. However, it is challenging for mMIMO systems to provide URLLC service for all devices, especially for the edge devices suffering from severe path loss or blockage \cite{ref20a}. Additionally, the inter-cell interference coming from adjacent cells becomes the bottleneck, when supporting the edge devices \cite{zhang2021joint}.

To address this issue, a user-centric cell-free mMIMO (CF mMIMO) network has been proposed in \cite{interdonato2019ubiquitous}. In contrast to the typical cell-centric networks, CF mMIMO can support user-centric transmissions, where a cluster of access points (APs) jointly serve a group of devices without strict cell boundaries \cite{ref16}. Owing to the effective collaboration among all geographically displaced APs, the desired signal power can be improved, while the inter-cell interference can be efficiently alleviated. Therefore, the capacity is increased accordingly \cite{ref14,2019ce}. Similar to the cooperation among base stations \cite{lozano2013fundamental}, the AP selection based on large-scale fading factors was proposed for solving the scalability issues in \cite{ref18b}. To reduce the implementation complexity, the distributed transmit precoding (TPC) schemes were investigated in \cite{ref18}. As a further development, the benefits of bespoke power control were studied in \cite{demir2020joint}. However, the existing methods are mainly based on Shannon's capacity under the assumption of infinite channel blocklength, which may not be applicable in short packet transmission. To fill this gap, only a few authors investigated the CF mMIMO-enabled URLLC \cite{CF21Per,ref20}. Explicitly, the authors of \cite{CF21Per} found that CF mMIMO can significantly increase the number of admitted devices requiring URLLC service over centralized mMIMO. \textcolor{black}{Furthermore, considerable data rate and energy efficiency improvements can be achieved using CF mMIMO in the FCBL regime \cite{ref20}}, where each AP was equipped with a single antenna. However, it is not possible to exploit the beneficial channel hardening feature via single-antenna APs  \cite{2018Channelharden}. With this in mind, the CF mMIMO-enabled URLLC concept relying on multiple-antenna APs is yet to be established.

Although CF mMIMO \textcolor{black}{systems are capable of providing} URLLC services for edge devices, challenges continue to arise. Due to the limited channel coherence time, it is impractical to assign unique orthogonal pilots to a large number of devices. Thus, each pilot sequence may be shared among several devices \cite{attarifar2018random}. This results in severe pilot contamination, which undoubtedly constitutes a performance bottleneck of CF mMIMO systems \cite{mai2018pilot}. To alleviate the interference caused by pilot reuse, it is important to judiciously allocate pilot patterns among the devices. Some researchers focused on minimizing the pilot length \cite{liu2020graph} or on minimizing pilot contamination, \textcolor{black}{by harnessing for example a} Tabu-search-based algorithm \cite{liu2019tabu}, the Hungarian algorithm \cite{buzzi2020pilot}, and the weighted graphic framework-based algorithm \cite{zeng2021pilot}. However, none of the existing pilot assignment strategies have considered the devices' requirements under finite channel blocklength.

To tackle the issue of supporting multiple devices, we aim for jointly optimizing the pilot length, pilot allocation, and power control in the CF mMIMO-aided URLLC. In general, there are two challenges in this system. Firstly, compared to conventional centralized mMIMO systems, it is more challenging to derive the ergodic data rate of CF mMIMO systems under short channel blocklength. Secondly, in contrast to conventional long packet transmission, it is challenging to determine the most appropriate blocklength for channel estimation and data transmission for guaranteeing multiple devices' rate and DEP requirements in the short packet transmission regime. Specifically, although allocating a higher blocklength used for channel estimation is capable of enhancing the accuracy of the estimated channel gain, it reduces the blocklength of payload transmission, hence resulting in the data rate degradation. Therefore, it is an open problem how to share a given blocklength between channel estimation and payload transmission as well as how to allocate the pilot to maximize the number of devices admitted in URLLC.

To address the above-mentioned challenges, our contributions are summarized as follows:
\begin{enumerate}
	\item By considering the impact of imperfect CSI and pilot contamination, we derive the LB of the downlink data rate under finite channel blocklength, which provides an explicit expression for our resource allocation design.
	\item As for the pilot reuse strategy, our objective is to support more admitted devices through judiciously designing the pilot allocation scheme, which is a computationally challenging problem. To address this, we first transform the pilot allocation problem into a minimum coloring problem. Then, considering the devices' requirements of DEP and data rate, we construct an undirected graph by connecting the devices that cannot share the common pilot sequence and then assign the pilot sequences having minimal pilot length. Finally, we propose a low-complexity algorithm for determining the pilot assignment strategy.
	\item The WSR is maximized by jointly optimizing the pilot power and the transmission power while additionally considering the minimal DEP and data rate requirements. To solve this NP-hard problem, we first transform these requirements into desired SINR. Then, by introducing approximations, the problem can be decomposed into a series of subproblems, which can be transformed into a geometric programming (GP) problem by using log-function approximation. Finally, an iterative algorithm is proposed for maximizing the WSR.
	\item Our simulation results demonstrate the significant improvement attained in terms of the number of admitted devices and the rapid convergence speed of the proposed algorithm. Furthermore, our simulation results also confirm its superiority over existing algorithms.
\end{enumerate}

The remainder of this paper is organized as follows. In Section II, the system model is provided, and the LB date rate expression under FCBL based on statistical CSI is derived for the downlink. By optimizing the pilot length and pilot allocation, the number of devices admitted is maximized in Section III. Then, the power allocation is optimized for maximizing the WSR in Section IV. Our simulation results are presented in Section V. Finally, our conclusions are offered in Section VI.

\section{System Model and Spectral Efficiency}
\subsection{System Model}
We consider a CF mMIMO-aided smart factory, where $M$ APs jointly serve all $K$ single-antenna devices. Besides, each AP is equipped with $N$ antennas. The channel vector ${\bf{g}}_{m,k} \in {\mathbb{C}}^{N \times 1} $ between the $m$th AP and the $k$th device is modeled as
\begin{equation}
\setlength\abovedisplayskip{5pt}
\setlength\belowdisplayskip{5pt}
\label{cofficient_g}
{{\bf{g}}_{m,k}} = \sqrt {{\beta _{m,k}}} {{\bf{h}}_{m,k}},
\end{equation}
where $\beta _{m,k}$ is the large-scale fading factor and ${\bf{h}}_{m,k} \in \mathcal{CN} \left( {{\bf{0}},{\bf{I}}_N} \right)$ represents the small-scale fading coefficient.

\subsection{Channel Estimation}
We assume that time division duplex (TDD) is adopted. All devices send their pilot sequences, and then the APs estimate the CSI relying on a finite channel blocklength of $L = B \times T_B$, where $B$ is the bandwidth and $T_B$ is the transmission duration. The channel blocklength $\tau$ is allocated for channel estimation and $(L - \tau)$ for data transmission. The received signal at the $m$th AP is given by
\begin{equation}
\label{received_pilot}
\setlength\abovedisplayskip{5pt}
\setlength\belowdisplayskip{5pt}
{{\bf{Y}}^p_{m}} = \sum\limits_{k = 1}^K {{{\bf{g}}_{m,k}}\sqrt {{\tau}p_k^p} {\bf{ q }}_k^H}  + {{\bf{N}}^p_{m}},
\end{equation}
where $p_k^p$ is the $k$th device's pilot power, ${\bf{q}}_k \in {\mathbb{C}}^{\tau \times 1}$ is the pilot sequence of the $k$th device, and ${{\bf{N}}_{m}^p} \in {{\mathbb{C}}^{N \times \tau}}$ is the additive Gaussian noise matrix, having independent elements that follow the distribution of $\mathcal{CN} \left( {{{0}},{{1}}} \right)$. By multiplying (\ref{received_pilot}) with the pilot ${\bf{ q }}_k$, we have
\begin{equation}
\label{received_channel}
\setlength\abovedisplayskip{5pt}
\setlength\belowdisplayskip{5pt}
{{\bf{\hat y}}_{m,k}^p} = \frac{1}{{\sqrt {\tau p_k^p} }}{{\bf{Y}}^p_{m}}{{\bf{q}}_k} = \sum\limits_{k' = 1}^K {\sqrt {\frac{{p_{k'}^p}}{{p_k^p}}} {{\bf{g}}_{m,k'}}{\bf{q}}_{k'}^H{{\bf{q}}_k} + } \frac{{{\bf{N}}_m^p{{\bf{q}}_k}}}{{\sqrt {\tau p_k^p} }}.
\end{equation}

According to (\ref{received_channel}), allocating orthogonal pilots for each device will avoid causing pilot contamination, yielding ${\bf{q}}_{k'}^H{{\bf{q}}_k}=0$, when $k'\neq k$. However, this would require long pilot sequences for assigning unique device with specific pilot, as the smart factory has to support a large number of devices. To reduce the pilot overhead, we consider sharing the pilots in our CF mMIMO-aided URLLC system amongst appropriately allotted devices, where the number of devices $K$ is higher than the blocklength $\tau$ allocated for channel estimation. Specifically, we define the set of devices that are allocated the $i$th pilot sequence as $\mathcal{Q}_i(i\in\{1,2,\cdots,\tau\})$. Based on these assumptions, the channel vector ${\bf{\hat g}}_{m,k}$ estimated by the minimum mean squared error (MMSE) method is
\begin{equation}
\label{estimate_gmk}
\setlength\abovedisplayskip{5pt}
\setlength\belowdisplayskip{5pt}
{{{\bf{\hat g}}}_{m,k}} = \frac{{\tau p_k^p{\beta _{m,k}}}}{{\sum\limits_{k' \in {\mathcal{Q}_k}} {\tau p_{k'}^p{\beta _{m,k'}}}  + 1}}{{{\bf{\hat y}}}^p_{m,k}},
\end{equation}
which follows the distribution of $\mathcal{CN} \left( {{{\bf {0}}},{{{\lambda_{m,k}}{{\bf{I}}_N}}}} \right)$, where $\lambda_{m,k}$ given by
\begin{equation}
\label{gama_mk}
\setlength\abovedisplayskip{5pt}
\setlength\belowdisplayskip{5pt}
{\lambda_{m,k}} = \frac{{\tau p_k^p{\beta _{m,k}^2}}}{{\sum\limits_{k' \in {\mathcal{Q}_k}} {\tau p_{k'}^p{\beta _{m,k'}}}  + 1}}.
\end{equation}
Then, the channel estimation error is ${{{\bf{\tilde g}}}_{m,k}} = {{\bf{g}}_{m,k}} - {{{\bf{\hat g}}}_{m,k}}$, which is independent of $ {{{\bf{\hat g}}}_{m,k}}$ and follows the distribution of $\mathcal{CN} \left( {{{\bf{0}}},{{ \left[ {{\beta _{m,k}} - {\lambda _{m,k}}} \right]{{\bf{I}}_N}}}} \right)$. 

\subsection{Downlink Data Transmission}
During data transmission, the user-centric approach is adopted for reducing the implementation complexity, where each AP only has to provide services for the nearby devices. Let us denote the set of APs that serve the $k$th device as ${\mathcal{M}}_k$ and  the set of devices that are served by the $m$th AP as ${\mathcal{U}}_m$, respectively.

Each AP relies on the estimated channel for constituting its maximum ratio transmission (MRT) precoding scheme to transmit its signals. \textcolor{black}{Based on the user-centric approach, the signal transmitted by the $m$th AP is given by
\begin{equation}
\label{transmitted_signal}
\setlength\abovedisplayskip{5pt}
\setlength\belowdisplayskip{5pt}
{{\bf{x}}_m} = \sum\limits_{k \in {{\cal U}_m}} {\sqrt {p_{m,k}^d} {{\bf {a}}_{m,k}^ *}{s_k}},
\end{equation}
where ${p_{m,k}^d}$ is the transmission power, ${{\bf{ a}}_{m,k}} = \frac{{{{{\bf{\hat g}}}_{m,k}}}}{{\sqrt {{\mathbb{E}}\left\{ {{{\left\| {{{{\bf{\hat g}}}_{m,k}}} \right\|}^2}} \right\}} }}$ is the TPC vector \cite{ref18,interdonato2018downlink}}, and $s_k$ is the data symbol transmitted to the $k$th device. Then, the signal received at the $k$th device is
\begin{align}
\setlength\abovedisplayskip{5pt}
\setlength\belowdisplayskip{5pt}
y_k^d &= \sum\limits_{m = 1}^M {\sum\limits_{k' \in {{\cal U}_m}} {\sqrt {p_{m,k'}^d} {\bf{g}}_{m,k}^T{\bf{a}}_{m,k'}^ * {s_{k'}}} }  + {n_k} \notag \\
& = \sum \limits_{k' = 1}^K {\sum\limits_{m \in {{\cal M}_{k'}}} {{{\left( {{{\bf{g}}_{m,k}}} \right)}^T}{\bf{a}}_{m,k'}^ * \sqrt {p_{m,k'}^d} {s_{k'}}} }  + {n_k}, \label{downlink_kth_signal}
\end{align}
where $n_k$ is the noise having the distribution of $\mathcal{CN} \left( {{{0}},{{1}}} \right)$. Without downlink pilots, the mean of the estimated channel gain is assumed to be the true channel for signal detection \cite{2018CSI}. As a result, the signal received by the $k$th device may be rewritten as
\begin{align}
\setlength\abovedisplayskip{5pt}
\setlength\belowdisplayskip{5pt}
y_k^d & = \underbrace { {\mathbb E}\left\{ {\sum\limits_{m \in {{\cal M}_k}} {{{\left( {{{\bf{\hat g}}_{m,k}}} \right)}^T}{\bf{a}}_{m,k}^ * \sqrt {p_{m,k}^d} } } \right\}}_{{{\rm{DS}}_k^d}}{s_k}  \notag \\
& + \underbrace {\left\{ {\sum\limits_{m \in {{\cal M}_k}} {{{\left( {{{\bf{g}}_{m,k}}} \right)}^T}{\bf{a}}_{m,k}^ * \sqrt {p_{m,k}^d} }  - {{{\rm{DS}}_k^d}}} \right\}}_{{\rm LS}_{k}^d}{s_k} , \label{downlink_kth_statistics_signal} \\
& + \sum\limits_{k' \ne k}^K {\underbrace {\sum\limits_{m \in {{\cal M}_{k'}}} {{{\left( {{{\bf{g}}_{m,k}}} \right)}^T}{{ {{\bf{a}}_{m,k'}^ * } } }\sqrt {p_{m,k'}^d} } }_{{\rm UI}_{{k,k'}}^d}{s_{k'}}}  + \underbrace {{n_k}}_{{{\rm N}_k^d}}, \notag
\end{align}
where ${\rm{DS}}_k^d$ is the desired signal, ${\rm {LS}}_k^d$ is the leaked signal, ${\rm {UI}}_{k,k'}^d$ represents the interference due to the $k'$th device, and ${\rm {N}}_k^d$ is the noise term. Finally, the SINR of the $k$th device is given by
\begin{equation}
\setlength\abovedisplayskip{5pt}
\setlength\belowdisplayskip{5pt}
\label{kth_SINR_downlink}
\gamma _k^d = \frac{{{{\left| {{\rm{DS}}_{k}^d} \right|}^2}}}{{{{\left| {{\rm{LS}}_{k}^d} \right|}^2} + \sum\nolimits_{k' \ne k}^K {{{\left| {{\rm{UI}}_{k,k'}^d} \right|}^2}}  + {{\left| {{{\rm{N}}_k^d}} \right|}^2}}}.
\end{equation}

\subsection{Achievable Data Rate under Finite Blocklength}
Again, the Shannon capacity \cite{1948Shanon} is defined as the maximum coding rate that may allow the DEP to approach zero, when the channel blocklength is infinity. However, in short packet transmissions, the DEP is usually significantly increased. \textcolor{black}{By treating the interference as a part of the Gaussian noise \cite{ref3,ref13d,li2020multi}, the achievable data rate of the $k$th device under FCBL can be approximated as}
\begin{equation}
\label{urllc_rate}
\setlength\abovedisplayskip{5pt}
\setlength\belowdisplayskip{5pt}
{R_k} \approx \left( {1 - \eta }\right) {\log _2}\left( {1 + \gamma_k} \right) - \sqrt {\frac{{\left( {1 - \eta } \right){V_k\left(\gamma_k\right)}}}{L}} \frac{{{Q^{ - 1}}\left( {{\varepsilon _k}} \right)}}{{\ln 2}},
\end{equation}
where $\eta = \tau / L$\footnote{Note that a pilot sequence may be shared among specific devices, and the pilot length $\tau$ is always smaller than the number of devices $ K$.}, ${\gamma}_k$ is the $k$th device's SINR, ${{\varepsilon _k}}$ is the DEP,  $V_k$ is the channel's dispersion associated with  ${V_k\left(\gamma_k\right)} = 1 - {\left( {1 + {\gamma_k}} \right)^{ - 2}}$, and ${Q^{ - 1}}\left( {{\varepsilon _k}} \right)$ is the inverse function of $Q\left( {{\varepsilon _k}} \right) = \frac{1}{{\sqrt {2\pi } }}\int_{{\varepsilon _k}}^\infty  {{{\rm{e}}^{{{ - {t^2}} \mathord{\left/
 {\vphantom {{ - {t^2}} 2}} \right.
 \kern-\nulldelimiterspace} 2}}}{\rm{d}}t}$ of the $k$th device.

The ergodic data rate of the $k$th device under FCBL is given by \cite{ref3}
\begin{equation}
\setlength\abovedisplayskip{5pt}
\setlength\belowdisplayskip{5pt}
\label{rw_rate}
\begin{split}
{\bar R_k} & \approx \mathbb{E} \left\{ \frac{{1 - \eta }}{{\ln 2}}\left[ {\ln \left( {1 + {\gamma _k}} \right) - \frac{{{ Q^{ - 1}}\left( {{\varepsilon _k}} \right)}}{{\sqrt {L\left( {1 - \eta } \right)} }}\sqrt {\frac{{\frac{2}{{{\gamma _k}}} + 1}}{{{{\left( {\frac{1}{{{\gamma _k}}} + 1} \right)}^2}}}} } \right] \right\}, \\
&\triangleq \frac{{1 - \eta }}{{\ln 2}} \mathbb{E} \left\{ f_k \left( \frac{1}{\gamma_k} \right)\right\},
\end{split}
\end{equation}
where $f_{k}(x) = \ln(1+x) - \frac{{{ Q^{ - 1}}\left( {{\varepsilon _k}} \right)}}{{\sqrt {L\left( {1 - \eta } \right)} }} \sqrt {\frac{{2x + 1}}{{{{\left( {x + 1} \right)}^2}}}}$ is a function in term of the $k$th device's DEP requirement. Furthermore, the expectation is taken over the small-scale fading factors. This is due to the fact that the packets should be delivered to the devices at extremely low latency, and thus large scale fading factors remain time-invariant. However, it is intractable to derive the closed-form expression of the ergodic data rate, and thus it is challenging to perform resource allocation directly based on (\ref{rw_rate}).

To address the above issue, we derive the LB of the ergodic data rate that facilitates resource allocation. Firstly, the data rate $R_k$ of any device cannot be lower than $0$, hence we have the following inequality
\begin{equation}
\setlength\abovedisplayskip{5pt}
\setlength\belowdisplayskip{5pt}
\label{a_region}
\frac{{{Q^{ - 1}}\left( {{\varepsilon _k}} \right)}}{{\sqrt {L\left( {1 - \eta } \right)} }} \le \frac{{\left( {\frac{1}{{{\gamma _k}}} + 1} \right)\ln \left( {1 + {\gamma _k}} \right)}}{{\sqrt {\frac{2}{{{\gamma _k}}} + 1} }} \buildrel \Delta \over =  g\left( \frac{1}{{\gamma}_k} \right ),
\end{equation}
where $g(x)$ is equal to $\frac{(x + 1)\ln(1 + \frac{1}{x})}{\sqrt{2x + 1}}$. It is readily to see that the first-order derivative of $g\left( x \right )$ is negative, which means that $g\left( x \right )$ is a monotonically decreasing function of $x$. Furthermore, the feasible region of $f_{k} \left(x \right)$ is $0 \le x \le {g^{ - 1}}\left( {\frac{{{Q^{ - 1}}\left( {{\varepsilon _k}} \right)}}{{\sqrt {L\left( {1 - \eta } \right)} }}} \right)$. As a result, we have the following lemma.

\begin{lemma}
\label{x_region}
Function $f_{k} \left(x \right)$ is a decreasing and convex function when $ 0 < x \le {g^{ - 1}}\left( {\frac{{{Q^{ - 1}}\left( {{\varepsilon _k}} \right)}}{{\sqrt {L\left( {1 - \eta } \right)} }}} \right) $. 

{\it Proof}: Please refer to Appendix B in \cite{ref20a}. $\hfill\blacksquare$
\end{lemma}

Using Lemma \ref{x_region} and Jensen's inequality, we have
\begin{equation}
\setlength\abovedisplayskip{5pt}
\setlength\belowdisplayskip{5pt}
\label{rate}
{\bar R_k} \ge {\hat R}_k \triangleq \frac{{1 - \eta }}{{\ln 2}}{f_k}\left( {{1 \mathord{\left/
 {\vphantom {1 {{{\hat \gamma }_k}}}} \right.
 \kern-\nulldelimiterspace} {{{\hat \gamma }_k}}}} \right),
\end{equation}
where ${\hat R}_k$ is the LB data rate of the $k$th device, and ${{\hat \gamma }_k}$ is ${{\hat \gamma }_k} = \frac{1}{\mathbb{E} \left( {{1 \mathord{\left/
 {\vphantom {1 {{{\hat \gamma }_k}}}} \right.
 \kern-\nulldelimiterspace} {{{ \gamma }_k}}}} \right)}$.

Based on the above-mentioned discussions, we have to derive ${{\hat \gamma }_k}$ in the remaining parts.

%
%
%

\begin{theorem}
	\label{LB_MRT_T}
	The ergodic data rate of the $k$th device using the FCBL MRT precoder can be lower bounded by
	\begin{equation}
		\setlength\abovedisplayskip{5pt}
		\setlength\belowdisplayskip{5pt}
		\label{MRT_LB_rate}
		{\hat R}_k^{d} \triangleq \frac{{1 - \eta }}{{\ln 2}} f_k \left( \frac{1}{{\hat \gamma }_k^{d}} \right),
	\end{equation}
	where ${\hat \gamma }_k^{d}$ is given by (\ref{MRT_SINR_LB}) at the bottom of this page.
\begin{figure*}[hb]
	\hrulefill
	\begin{equation}
		\setlength\abovedisplayskip{5pt}
		\setlength\belowdisplayskip{5pt}
		\label{MRT_SINR_LB}
		 \hat \gamma _k^{d} = \frac{{{{\left( {\sum\limits_{m \in {{\cal M}_k}} {\sqrt {Np_{m,k}^d{\lambda _{m,k}}} } } \right)}^2}}}{{\sum\limits_{k' = 1}^K {\sum\limits_{m \in {{\cal M}_{k'}}} {p_{m,k'}^d{\beta _{m,k}}} }  + N\sum\limits_{k' \in \left\{ {{{\cal Q}_k}\backslash k} \right\}} {{{\left( {\sum\limits_{m \in {{\cal M}_{k'}}} {\sqrt {p_{m,k'}^d{\lambda _{m,k}}}} } \right)}^2}}  + 1}}.
	\end{equation}
\end{figure*}
	
	\emph{Proof}: Please refer to Appendix \ref{Proof_MRT_LB}. $\hfill\blacksquare$
	
\end{theorem}

Based on (\ref{MRT_LB_rate}), the LB data rate is determined by the pilot power, payload power, and pilot allocation strategy, which motivates us to appropriately allocate the resources to guarantee the devices' rate and DEP requirements while relying on a finite channel blocklength.

\section{Pilot Assignment}
Based on the downlink data rate LB in (\ref{MRT_LB_rate}), we aim for maximizing the number of admitted devices by jointly optimizing the pilot length and the pilot allocation.

\subsection{Problem Formulation}
Traditionally, the devices served by AP sets that have at least one AP in common should be assigned unique orthogonal pilot sequences, which can be formulated as \cite{biguesh2006training}
\begin{equation}
	\setlength\abovedisplayskip{5pt}
	\setlength\belowdisplayskip{5pt}
	\label{orthogonal_mth_AP}
	{\bf{q}}_k^H{\bf{q}}_{k'} = 0, \quad  {\text{for}} \quad {{{\mathcal{M}}_k} \cap {{\mathcal{M}}_{k'}} \ne \emptyset, {\text{and}} \quad k \ne k'}.
\end{equation}
For ease of exposition, we define a $K \times K$ binary matrix $\bf B$ to indicate whether the $k$th device and the $k'$th device are allowed to share the same pilot pattern. The $k$th row and $k'$th column element of the matrix $\bf B$ is given by
\begin{equation}
	\setlength\abovedisplayskip{5pt}
	\setlength\belowdisplayskip{5pt}
	\label{bkk}
	{b_{k,k'}} = \left\{ {\begin{array}{*{20}{c}}
			{1,}&{{{\mathcal{M}}_k} \cap {{\mathcal{M}}_{k'}} \ne \emptyset, {\text{and}} \quad k \ne k'}\\
			{0,}&{{\text{otherwise,}}}
	\end{array}} \right.
\end{equation}
where $b_{k,k'} = 1$ means that the $k$th device and the $k'$th device are served by AP sets that have at least one AP in common. Therefore, these devices should be allocated unique orthogonal pilot sequences. By contrast, for $b_{k,k'} = 0$, it indicates that the $k'$th device is a potential candidate for sharing the pilot sequence with the $k$th device.

To allocate the channel estimation resources in a fair manner, the maximum reuse time for all pilot sequences is defined as $n_{\text{max}}$. As a result, the number of devices in set $\mathcal{Q}_i$ should be no higher than $n_{\text{max}}$, which can be expressed as
\begin{equation}
	\setlength\abovedisplayskip{5pt}
	\setlength\belowdisplayskip{5pt}
	\label{fair_manner}
	 |\mathcal{Q}_i| \le n_{\text{max}}, \forall i \in \left\lbrace 1,2, \cdot \cdot \cdot, \tau \right\rbrace.
\end{equation}
This constraint can prevent the worst-case scenario where a single pilot sequence is reused by too many devices.

We aim for striking a trade-off between the pilot blocklength $\tau$ used for channel estimation and the data transmission blocklength $(L-\tau)$ whilst maximizing the number of devices admitted. Based on the LB data rate derived, we denote the set of devices admitted that satisfy the requirements of DEP, latency (blocklength), and data rate constraints as ${\mathcal{S}} = \left\{ {k\left| {{{\hat R}_k^{d}} \ge R_k^{\text{req}},\forall k} \right.} \right\}$, where ${{\hat R}_k^{d}}$ and $R_k^{\text{req}}$ are the $k$th device's data rate LB and the required data rate, respectively. Since many factors affect the LB rate ${{\hat R}_k^{d}}$, we fix the pilot power and payload transmission power to investigate the impact of pilot length and pilot allocation on the data rate attained. Essentially, we assume that the $k$th device adopts the maximum power to transmit its pilot sequence (i.e. $p_k^p = P_k^{\rm max,p}$, $\forall k$) and that the $m$th AP equally shares the downlink transmission power of the devices in the set of ${\cal{U}}_m$, which is detailed as follows
\begin{equation}
	\setlength\abovedisplayskip{5pt}
	\setlength\belowdisplayskip{5pt}
	\label{fair_downlink_power}
	{p_{m,k}^d} = \left\{ {\begin{array}{*{20}{c}}
			{\frac{{{P_m}}}{{\left| {{\cal{U}}_m} \right|}},}&{k \in {\cal{U}}_m}\\
			{0,}&{k \notin {\cal{U}}_m}
	\end{array}} \right.,
\end{equation} 
where $P_m$ is the maximum transmission power of the $m$th AP. Then, by substituting the power allocation and pilot strategy into (\ref{MRT_LB_rate}), we can readily obtain the LB rate ${{\hat R}_k^{d}}$. Finally, by optimizing the blocklength $\tau$ and by appropriately allocating the pilot sequences, we aim for maximizing the number of devices in $\mathcal{S}$, which is formulated as
\begin{equation}
	\setlength\abovedisplayskip{5pt}
	\setlength\belowdisplayskip{5pt}
	\label{Pilot_assignment}
	\begin{array}{*{20}{c}}
		{\mathop {\max }\limits_{\left\{ {{\mathcal Q}_i} \right\},\tau} }&{{|\mathcal{S}|}}\\
		{{\rm{s}}{\rm{.t}}{\rm{.}}}& (\text{\ref{orthogonal_mth_AP}}),(\text{\ref{fair_manner}}).
	\end{array}
\end{equation}
 For solving Problem (\ref{Pilot_assignment}), the computational complexity grows exponentially both with the number of devices and the number of pilot sequences. In the following, we conceive a low-complexity pilot allocation scheme, which is applicable to practical systems.

\subsection{Pilot Allocation Scheme}

\begin{figure*}
	\centering
	\subfigure[\textcolor{black}{Undirected graph based on $\bf{B}$.}]{
		\includegraphics[width=2.75in]{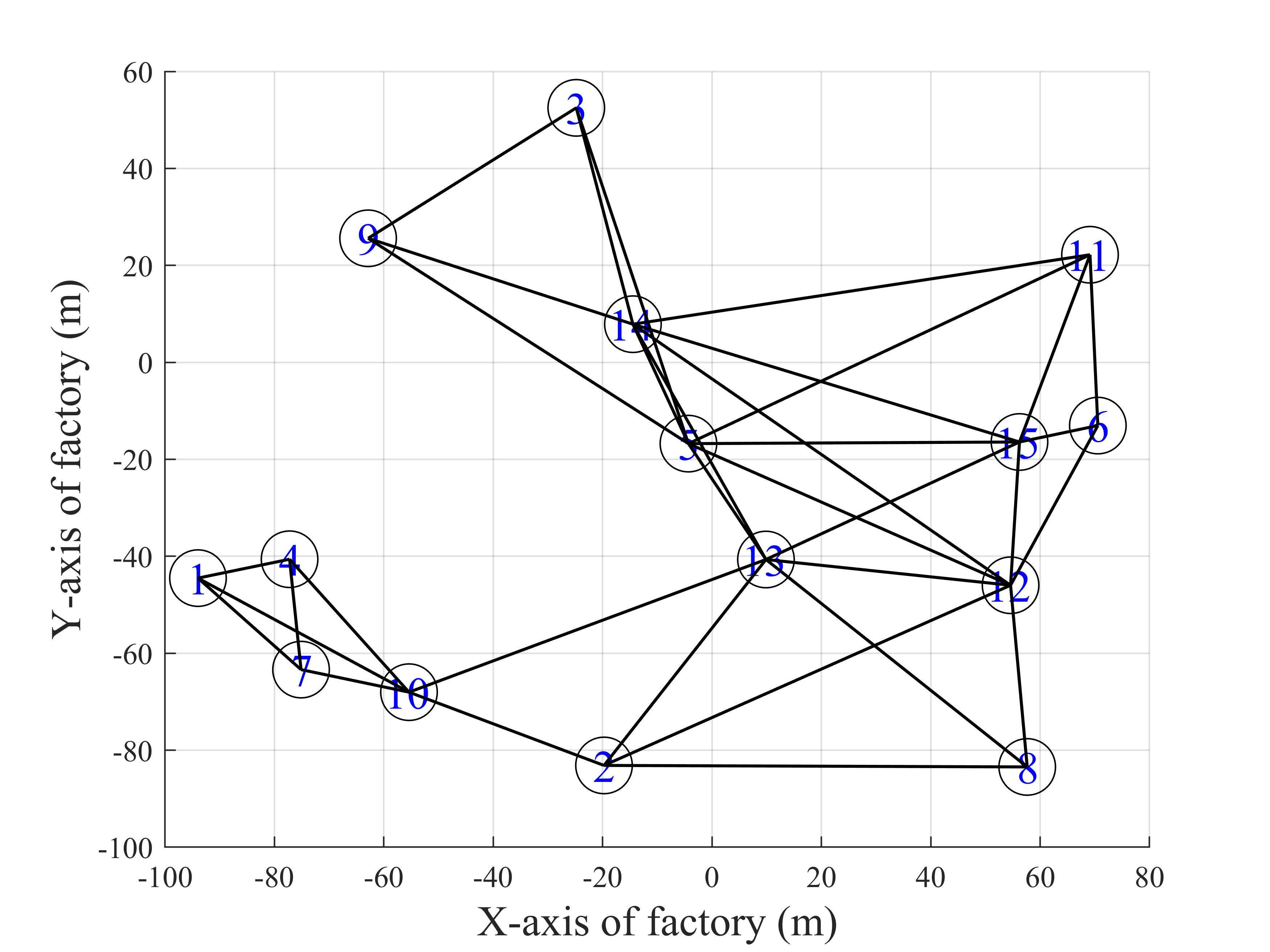}}\hspace{10mm}
	\subfigure[\textcolor{black}{Pilot assignment based on the Dsatur algorithm.}]{
		\includegraphics[width=2.75in]{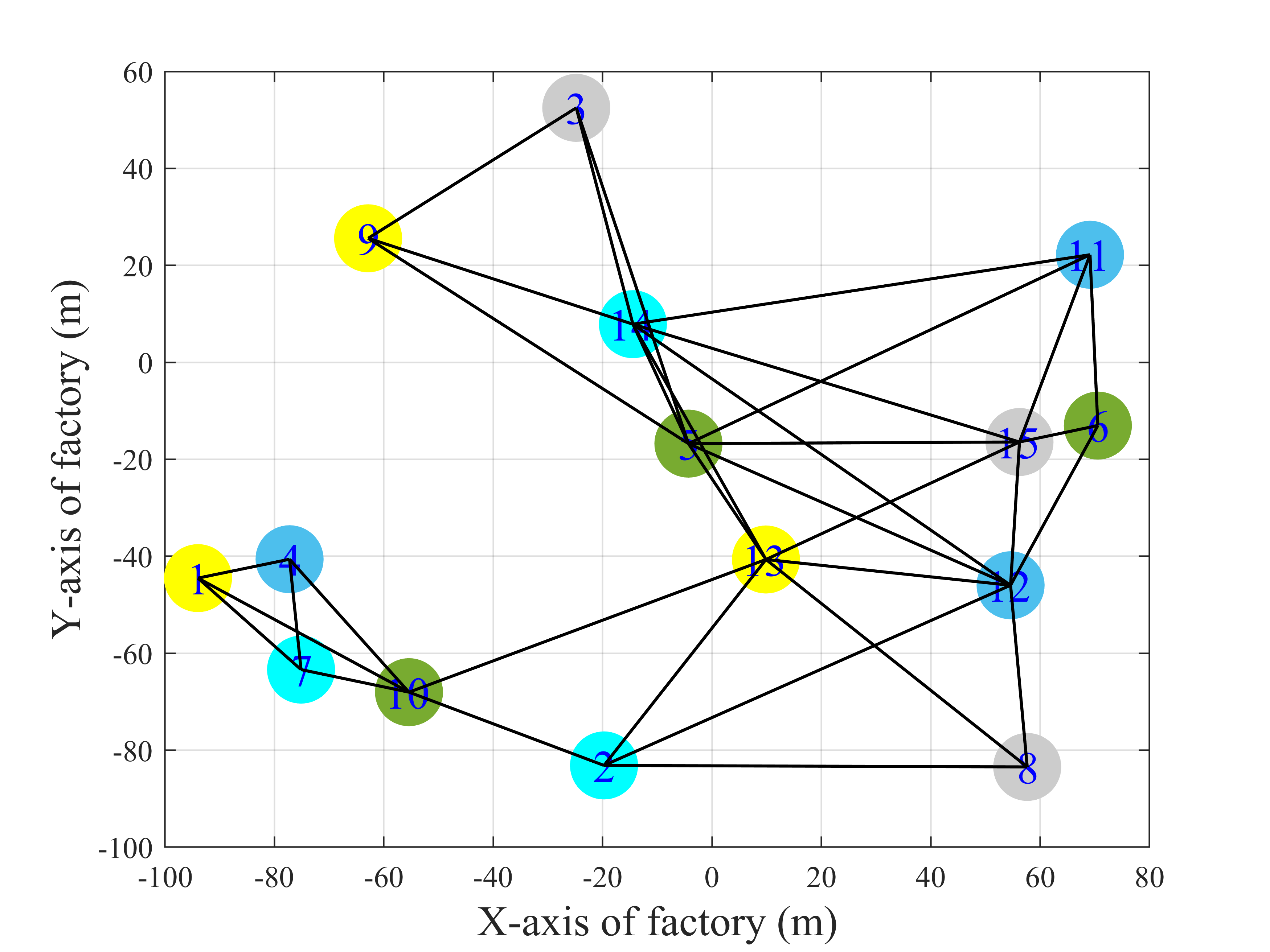}}
	
	\subfigure[\textcolor{black}{Undirected graph by using proposed pilot strategy}.]{
		\includegraphics[width=2.75in]{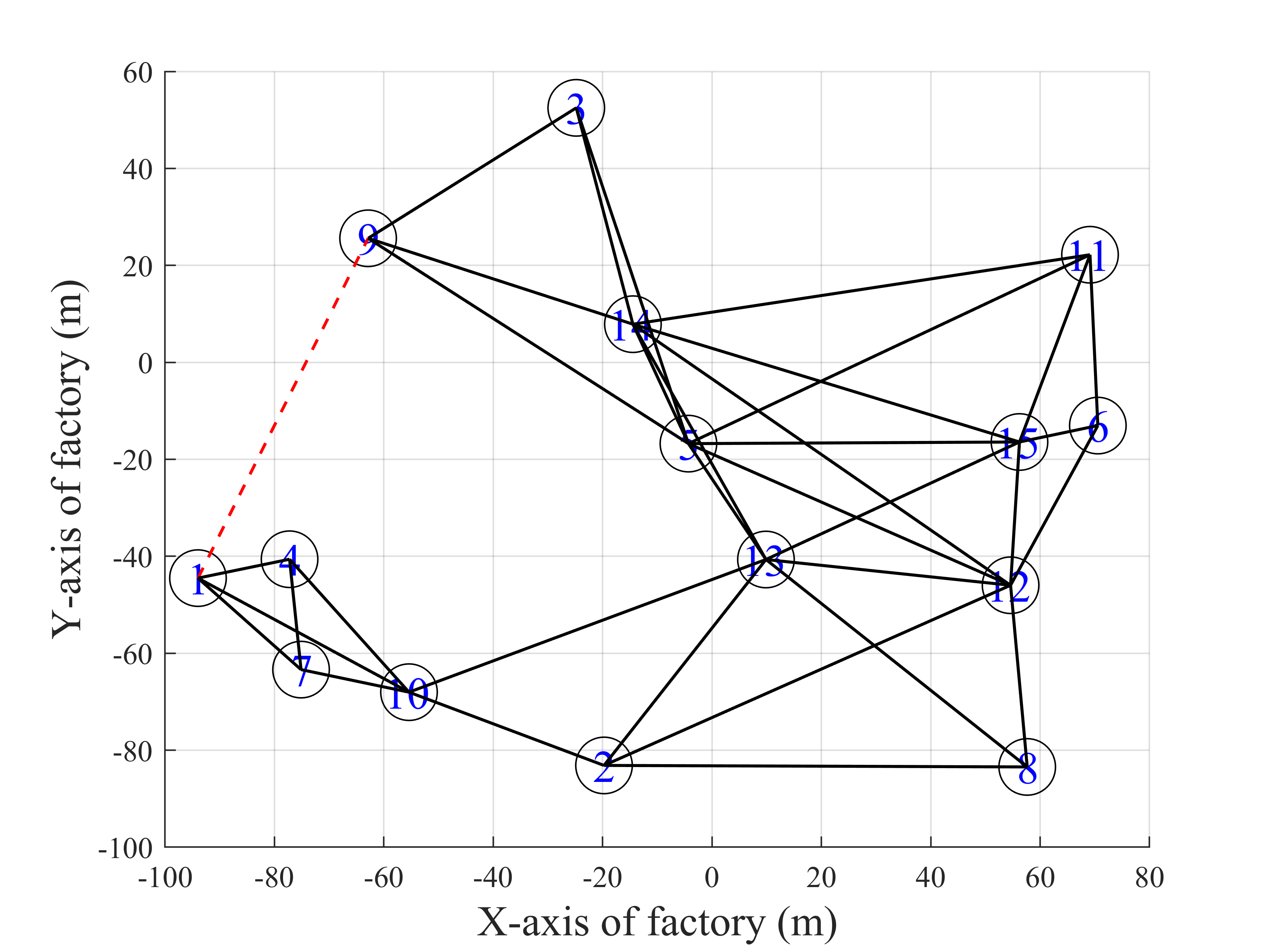}}\hspace{10mm}
	\subfigure[\textcolor{black}{Pilot allocation based on the proposed algorithm.}]{
		\includegraphics[width=2.75in]{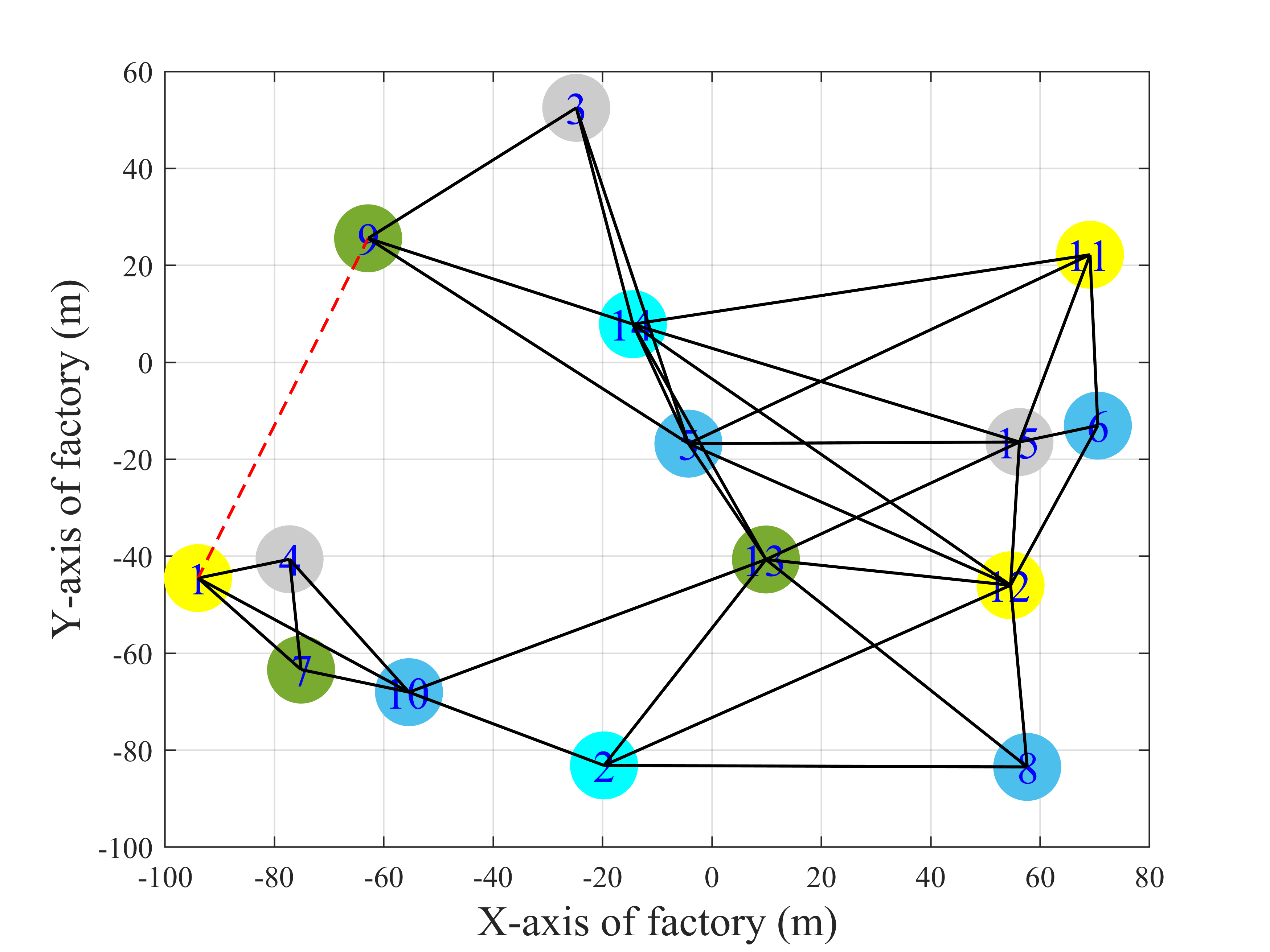}}
	\centering
	\caption{\textcolor{black}{Undirected graph and pilot assignment based on the Dsatur and proposed algorithms.}}
	\label{fig2}
\end{figure*}

To explore whether the devices can share the same pilot, an undirected graph based on matrix $\bf{B}$ is constructed, as illustrated in Fig. \ref{fig2}(a). The circle with number $k$ represents the $k$th device's location. Furthermore, if $b_{k,k'}$ is equal to 1, then the $k$th device and the $k'$th device are connected, where these two devices cannot share the same pilot sequence. By contrast, the unconnected devices imply that it is possible to share the same pilot sequence among these devices. Then, the pilot assignment problem can be transferred into a graph coloring problem. This minimal coloring scheme can be readily obtained by the Dsatur algorithm \cite{brelaz1979new} and the resultant solution is illustrated in Fig. \ref{fig2}(b), where the same color indicates that the same pilot sequence is used. Furthermore, it can be seen from Fig. \ref{fig2}(a) that only 5 pilot sequences are needed for supporting as many as 15 devices, thereby increasing the fraction of blocklength available for payload data transmission. It is worth noting that the Dsatur algorithm-based pilot assignment only aims for reducing the pilot overhead without considering the DEP, latency, and data rate requirements. To address this, we conceive a low-complexity algorithm for pilot allocation.



Let us denote the Dsatur algorithm-based pilot allocation strategy as $\mathfrak{L}_{\rm{Dsa}}\triangleq\{{\mathcal{Q}}_i^{\rm{Dsa}}, i= 1,2, \cdots, \tau_{\rm{Dsa}}\}$, where $\tau_{\rm{Dsa}}$ is the number of pilot sequences and ${\mathcal{Q}}_i^{\rm{Dsa}} \subset \{1,2,\cdots,K\}$ is the set of devices that share the $i$th pilot sequence, which can be obtained by using the Dsatur algorithm\footnote{According to \cite{brelaz1979new}, each device only uses a single sequence so that ${\mathcal{Q}}_i^{\rm{Dsa}}\bigcap {\mathcal{Q}}_j^{\rm{Dsa}}= \emptyset$, $\forall i\neq j$, holds.}. Obviously, the strategy can satisfy constraints (\ref{orthogonal_mth_AP}) and (\ref{fair_manner}), but may not maximize the number of devices admitted. To check whether this scheme can meet all the devices' QoS, we have to calculate the set of devices admitted, which is formulated as $\mathcal{S}_{\rm{Dsa}} = \left\lbrace k| {\hat R}_{k }^{d,(\mathfrak{L}_{\rm Dsa})} \ge R_k^{\rm{req}}, \forall k \right\rbrace$, where ${\hat R}_{k }^{d,(\mathfrak{L}_{\rm Dsa})}$ is obtained by substituting the pilot allocation scheme into (\ref{MRT_LB_rate}). Then, by substituting the pilot allocation strategy, the pilot power, and the transmission power into the downlink data rate LB in (\ref{MRT_LB_rate}), we can readily obtain the set of admitted devices $\mathcal{S}_{\rm{Dsa}}$. \textcolor{black}{If $|\mathcal{S}_{\rm{Dsa}}| = K$, then the current strategy $\mathfrak{L}_{\rm{Dsa}}$ based on the Dsatur algorithm can satisfy all the devices' QoS targets. By contrast, if $|\mathcal{S}_{\rm{Dsa}}|< K$, then the Dsatur algorithm-based pilot strategy cannot meet all the devices' QoS targets, hence a more appropriate solution of Problem (\ref{Pilot_assignment}) should be found.} 


For the case of $|\mathcal{S}_{\rm Dsa}|< K$, some devices should not share the pilot sequence with some particular devices. As a result, we propose a low-complexity iterative algorithm for finding those devices. For ease of exposition, we denote the pilot allocation strategy in the $r$th iteration as $\mathfrak{L}_{\rm{pro}}^{(r)}\triangleq\{{\mathcal{Q}}_i^{{\rm {pro}}(r)},  i =  1,2, \cdots \tau^{(r)}_{\rm pro}\}$, the binary matrix in the $r$th iteration as ${\bf{B}}^{(r)}$, and the maximum number of devices admitted throughout all $r$ iterations as $a^{(r)}$. The set of devices that fail to satisfy the QoS requirement when using the $i$th pilot sequence in the $r$th iteration is denoted as ${\mathcal{W}}_i^{(r)} = \left\{k' | {\hat R}_{k'}^{d,(\mathfrak{L}_{\rm{pro}}^{(r)})} - R_{k'}^{\rm req} < 0,\forall k' \in {\mathcal{Q}}_i^{{\rm{pro}}(r)} \right\}$,  $\forall i \in \{ 1,2, \cdot \cdot \cdot, \tau^{(r)}_{\rm pro} \}$, and the most unacceptable device in the set ${\mathcal{W}}_i^{(r)}$ is defined as $d^{(r)}_i = \arg \mathop {\min }\limits_{j\in \mathcal{W}_i^{(r)}} \left\{ { {\hat R}_{j}^{d,(\mathfrak{L}_{\rm{pro}}^{(r)})} - R_{j}^{{\rm{req}}}} \right\}$. Furthermore, the device that infects the most severe interference upon the $d^{(r)}_i$th device is defined as $c^{(r)}_i$ with $ c^{(r)}_i\in \left\{ {{\cal Q}_i^{{\rm{pro}}(r)}\backslash d_i^{(r)}} \right\}$. Since each device using the $i$th pilot sequence imposes pilot contamination on the $d^{(r)}_i$th device, we can find the $c^{(r)}_i$th device by finding the maximum interference induced by pilot contamination. Specifically, we first adopt the maximum pilot power $P_k^{\max,p}$ to calculate the estimated channel gain $\lambda_{m,k}, \forall m,k$, and then initialize the downlink transmission power by using (\ref{fair_downlink_power}). Then, the $c^{(r)}_i$th device can be obtained by $c^{(r)}_i = \arg \mathop {\max }\limits_{j \in \left\{ {{\cal Q}_i^{{\rm{pro}}\left( r \right)}\backslash d_i^{(r)}} \right\}} {\sum\limits_{m \in {{\cal M}_{j}}} {\sqrt {p_{m,j}^d{\lambda _{m,d^{(r)}_i}}}}}$. To mitigate the pilot contamination, we update the matrix ${\bf{B}}^{(r)}$ by setting ${b_{d^{(r)}_i,c_i^{\left( t \right)}}}$ and ${b_{c_i^{\left( t \right)},d^{(r)}_i}}$ as 1. Finally, we update the iteration $r$ as $r = r + 1$ and obtain the matrix ${\bf B}^{(r)}$ in the $r$th iteration by ${\bf B}^{(r)} = {\bf B}^{(r-1)}$. Based on the updated ${\bf{B}}^{(r)}$, we can now construct a new undirected graph and reassign the pilot sequences by using the Dsatur algorithm. Furthermore, by connecting the unallowed devices, we require an extended blocklength on channel estimation, since the pilot length $\tau_{\rm pro}^{(r)}$ may be increased with the number of iterations. To strike a performance\footnote{By adopting a rational pilot allocation scheme and reducing pilot overhead, more blocklength can be utilized for data transmission so that the QoS of devices can be satisfied as much as possible.} vs. implementation complexity trade-off, we define the maximum number of iterations as $N_{\max}^{\rm iter}$ and a search range for the pilot length as $(\tau_{\rm Dsa} + \iota)$ where $\iota$ is the extra search region of the extended pilot length. If the number of iterations or the search region of the pilot length exceeds the maximum tolerate value, we will output the best results found as the final pilot allocation strategy $\mathfrak{L}_{\rm{pro}}\triangleq\{{\mathcal{Q}}_i^{\rm pro},  i =  1,2, \cdots \tau_{\rm pro}\}$. The iterative procedure is detailed in Algorithm \ref{Pilot_Assignment}.

By harnessing our proposed approach, the undirected graph of Fig. \ref{fig2}(a) can be transformed into Fig. \ref{fig2}(c) by connecting the devices that must not share a common pilot. By taking the devices' rate and DEP requirements into consideration, it can be observed that the first device cannot share the pilot sequence with the devices linked by dotted lines. Hence, the pilot patterns in Fig. \ref{fig2}(b) cannot guarantee the devices' QoS. Instead, the pilot allocation strategy is rearranged in a more appropriate way, as illustrated in Fig. \ref{fig2}(d). Furthermore, compared to the pilot allocation seen in Fig. \ref{fig2}(b), the proposed approach significantly increases the minimal distance between devices that share a common pilot sequence, hence reducing the interference caused by pilot contamination.

\textcolor{black}{\subsection{Complexity Analysis}
	In this subsection, we provide the complexity analysis for the proposed pilot assignment. The complexity of the pilot allocation scheme depends on the product of the number of iterations and the complexity of each iteration. Specifically, the computational complexity of the Dsatur algorithm is on the order of $\mathcal{O}(K^2)$ [37], and thus the complexity of our pilot assignment is on the order of ${\mathcal{O}}(N_{it}K^2)$, where $N_{it}$ is the number of iterations with $1 < N_{it} \le N_{\max}^{\rm iter}$.}

\begin{algorithm}[t]
	\caption{Iterative Algorithm for Pilot Assignment}
	\begin{algorithmic}[1]
		\label{Pilot_Assignment}
		\STATE Construct the undirected graph based on $\bf B$, and adopt the Dsatur algorithm to obtain the pilot scheme $\mathfrak{L}_{\rm{Dsa}}  = \{{\mathcal{Q}}_i^{{\rm {Dsa}}},  i =  1,2, \cdots \tau_{\rm Dsa}\}$ with pilot length $\tau_{\rm{Dsa}}$;
		\STATE Initialize the uplink pilot power $p_k^p$ to be $P_k^{p,\rm max}$ W and downlink transmission power by using (\ref{fair_downlink_power}), and calculate the set of admitted devices $\mathcal{S}_{\rm{Dsa}} = \left\lbrace k| {\hat R}_{k }^{d,(\mathfrak{L}_{\rm Dsa})} \ge R_k^{\rm{req}}, \forall k \right\rbrace$;
		\IF{($|\mathcal{S}_{\rm{Dsa}}| < K$)}
		\STATE Initialize $\mathfrak{L}_{\rm{pro}}=\mathfrak{L}_{\rm{Dsa}}$, $r = 0$, $a^{(r)} = |{\mathcal S}_{\rm{Dsa}}|$, ${\bf{B}}^{(r)} = \bf{B}$,  $\mathfrak{L}_{\rm{pro}}^{(r)}=\mathfrak{L}_{\rm{Dsa}}$, ${\mathcal{Q}}_i^{{\rm {pro}}(r)}={\mathcal{Q}}_i^{{\rm {Dsa}}}$, and $\tau^{(r)}_{\rm pro}=\tau_{\rm{Dsa}}$;
		\WHILE{($\tau^{(r)}_{\rm {pro}} - \tau_{\rm{Dsa}} < \iota$ and $r < N_{\max}^{\rm iter}$)}
		\STATE Find the set of unallowed devices that utilize the $i$th pilot sequence via ${\mathcal{W}}_i^{(r)} = \left\{k' | {\hat R}_{k'}^{d,(\mathfrak{L}_{\rm{pro}}^{(r)})} - R_{k'}^{\rm req} < 0,\forall k' \in {\mathcal{Q}}_i^{{\rm {pro}}(r)}  \right\}$,  $\forall i \in \{ 1,2, \cdot \cdot \cdot, \tau^{(r)}_{\rm pro} \}$, using $d^{(r)}_i = \arg \mathop {\min }\limits_{j \in {\mathcal{W}}_i^{(r)}} \left\{ {{\hat R}_{j}^{d,(\mathfrak{L}_{\rm{pro}}^{(r)})} - R_{j}^{{\rm{req}}}} \right\}$ to search the most unacceptable device $d^{(r)}_i$, find the $c_i^{\left( t \right)}$th device that incurs the largest interference to the $d^{(r)}_i$th device, and update ${\bf{B}}^{(r)}$ with ${b_{d^{(r)}_i,c_i^{\left( t \right)}}} = {b_{c_i^{\left( t \right)},d^{(r)}_i}} = 1$;
		\STATE Update $r$ as $r = r + 1$, obtain the matrix ${\bf{B}}^{(r)}$ by ${\bf{B}}^{(r)} = {\bf{B}}^{(r - 1)}$, construct an undirected graph based on ${\bf{B}}^{(r)}$, adopt the Dsatur algorithm to obtain the pilot assignment $\mathfrak{L}_{\rm{pro}}^{(r)}$ and pilot length $\tau^{(r)}_{\rm pro}$, and calculate the set of admitted devices in the $r$th iteration $\mathcal{S}^{(r)} = \left\lbrace k| {\hat R}_{k}^{d,(\mathfrak{L}_{\rm{pro}}^{(r)})} \ge R_k^{\rm{req}}, \forall k \right\rbrace$;
		\IF{$(|\mathcal{S}^{(r)}| > a^{(r-1)})$}
			\STATE Update $a^{(r)} = |\mathcal{S}^{(r)}|$ and pilot allocation strategy $\mathfrak{L}_{\rm{pro}} = \{{{\cal{Q}}_i^{{\rm pro}(r)}}, \forall i = 1,2,\cdot \cdot \cdot,\tau^{(r)}_{\rm pro}\}$;
			\ELSE
				\STATE  Update $a^{(r)} = a^{(r-1)}$;
			\ENDIF
		\ENDWHILE
		\ENDIF

	\end{algorithmic}
\end{algorithm}


\section{Power Allocation}
In this section, \textcolor{black}{since the WSR is key performance indicator (KPI) for satisfying various devices' requirements, we aim for maximizing the WSR of CF mMIMO by jointly optimizing the pilot and payload data power allocation}.
\subsection{Problem formulation}
The WSR is maximized while considering the finite energy constraints of the devices and APs, as well as the devices' data rate requirement. Specifically, the problem can be formulated as
\begin{subequations}
	\setlength\abovedisplayskip{5pt}
	\setlength\belowdisplayskip{5pt}
	\label{downlink_optimization}
	\begin{align}
		\mathop {\max }\limits_{\left\{ {p_k^p} \right\},\left\{ {p_{m,k}^d} \right\}} & \sum\limits_{k = 1}^K {{w_k}{{\hat R}_k^{d}} } \label{downlink_obj}\\
		{\rm{s}}{\rm{.t}}{\rm{.}}\;\;\;\; & {{\hat R}_k^{d} } \ge R_k^{{\rm{req}}},\forall k,  \label{downlink_req_data} \\
		& {p_k^p \le P_k^{\max ,p}},\forall k \label{downlink_pilot_power}\\
		& {\sum\limits_{k \in {{\cal U}_m}}p_{m,k}^d \le {P_m},\forall m}, \label{downlink_mth_AP_power}
	\end{align}
\end{subequations}
where $w_k$, $R_k^{\rm{req}}$, and $P_k^{\max ,p}$ represent the $k$th device's weight, the minimal data rate requirement, and the maximal transmission power, respectively. Constraint (\ref{downlink_req_data}) indicates that all devices should satisfy the data rate requirements, constraint (\ref{downlink_pilot_power}) means that the $k$th device's uplink pilot power is limited, and constraint (\ref{downlink_mth_AP_power}) is the $m$th AP's total payload transmission power. 

Compared to the widely studied max-min optimization problem, it is more challenging to obtain an optimal solution to Problem (\ref{downlink_optimization}) as the WSR maximization problem is an NP-hard problem. Furthermore, \textcolor{black}{given} the complex data rate expression of (\ref{urllc_rate}), this optimization problem is challenging to solve. To tackle this, we first simplify the problem.

By using Lemma \ref{x_region}, the constraint (\ref{downlink_req_data}) can be simplified into the $k$th device's minimal SINR requirement, formulated as
\begin{equation}
	\setlength\abovedisplayskip{5pt}
	\setlength\belowdisplayskip{5pt}
	\label{transformation}
	\hat \gamma _k^{d} \ge \frac{1}{{f_k^{ - 1}\left( {\frac{{R_k^{{\rm{req}}}\ln 2}}{{(1 - \eta) }}} \right)}}.
\end{equation}
Then, by introducing the auxiliary variables ${\chi _k}$, Problem  (\ref{downlink_optimization}) can be equivalently transformed into the following optimization problem
\begin{subequations}
	\setlength\abovedisplayskip{5pt}
	\setlength\belowdisplayskip{5pt}
	\label{downlink_optimization_trans}
	\begin{align}
		\mathop {\max }\limits_{\left\{ {p_k^p} \right\},\left\{ {p_{m,k}^d} \right\},\left\{ {\chi_k} \right\}} & {\sum\limits_{k = 1}^K {{w_k}\frac{{\left( {1 - \eta } \right)}}{{\ln 2}}\left[ {\ln \left( {1 + {\chi _k}} \right) - {\alpha _k}G\left( {{\chi _k}} \right)} \right]} } \label{downlink_obj_trans_a}\\
		\text{s.t.} \;\;\;\;\; & \hat \gamma _k^{d} \ge {\chi _k},\forall k,  \label{downlink_obj_trans_b}\\
		&{\chi _k} \ge \frac{1}{{f_k^{ - 1}\left( {\frac{{R_k^{{\rm{req}}}\ln 2}}{{(1 - \eta) }}} \right)}},\forall k, \label{downlink_obj_trans_c} \\
		&{\rm{ \left(\ref{downlink_mth_AP_power}\right)}}, \label{downlink_obj_trans_d}
	\end{align}
\end{subequations}
where $G\left( {{\chi _k}} \right)$ is defined as $G\left( {{\chi _k}} \right) = \sqrt {1 - {{\left( {1 + {\chi _k}} \right)}^{ - 2}}}$, and $\alpha_k$ is $\alpha_k = {\frac{{{Q^{ - 1}}\left( {{\varepsilon _k}} \right)}}{{\sqrt {L\left( {1 - \eta } \right)} }}}$.

As seen in (\ref{downlink_obj_trans_a}), the objective function is a complex expression due to $G\left(\chi_k\right)$. Based on Lemma 3 and Lemma 4 of \cite{ref13d}, $\ln \left( {1 + {\chi _k}} \right) $ and $G\left( {{\chi _k}} \right)$ in (\ref{downlink_obj_trans_a}) can be approximated by the log-function approximation method. As a result, the WSR can be lower bounded in an iterative manner, which is detailed as follows
\begin{align}
	\setlength\abovedisplayskip{5pt}
	\setlength\belowdisplayskip{5pt}
	&{{w_k}\frac{{\left( {1 - \eta } \right)}}{{\ln 2}}\left[ {\ln \left( {1 + {\chi _k}} \right) - {\alpha _k}G\left( {{\chi _k}} \right)} \right]} \\	\label{downlink_equal_OF}
	\ge &{{w_k}\frac{{\left( {1 - \eta } \right)}}{{\ln 2}}\left[ {\ln {{\left( {{\chi _k}} \right)}^{\left[ {{\rho ^{\left( i \right)}_{k}} - {\alpha _k}{{\hat \rho }^{\left( i \right)}_{k}}} \right]}} + {\delta ^{\left( i \right)}_{k}} - {\alpha _k}{{\hat \delta }^{\left( i \right)}_{k}}} \right]} \notag,
\end{align}
where ${\rho ^{\left( i \right)}_{k}}$ is $\frac{{\chi}_k^{\left( i \right)}}{{1 + {\chi}_k^{\left( i \right)}}}$, ${\delta ^{\left( i \right)}_{k}}$ is $\left[ \ln \left( {1 + {\chi}_k^{\left( i \right)}} \right) - {\rho ^{\left( i \right)}_{k}}\ln \left( {{\chi}_k^{\left( i \right)}} \right)\right] $, ${{\hat \rho }^{\left( i \right)}_{k}}$ is $\left[ \frac{{{\chi}_k^{\left( i \right)}}}{{\sqrt {{{{\chi}_k^{\left( i \right)}}^2} + 2{\chi}_k^{\left( i \right)}} }} - \frac{{{\chi}_k^{\left( i \right)}\sqrt {{{{\chi}_k^{\left( i \right)}}^2} + 2{\chi}_k^{\left( i \right)}} }}{{{{\left( {1 + \chi_k^{\left(i\right)}} \right)}^2}}}\right] $, and ${{\hat \delta }^{\left( i \right)}_{k}}$ is $ \left[ \sqrt {1 - \frac{1}{{{{\left( {1 +{\chi}_k^{\left( i \right)}} \right)}^2}}}}  - {{\hat \rho }^{\left( i \right)}_{k}} \ln \left( {{\chi}_k^{\left( i \right)}} \right)\right] $ in the $i$th iteration. Furthermore, the equality only holds when $\chi_k = \chi_k^{\left(i \right)}$. \textcolor{black}{As a result, the WSR can be lower-bounded by
	\begin{equation}
		\setlength\abovedisplayskip{5pt}
		\setlength\belowdisplayskip{5pt}
		\label{downlink_WSR}
		\begin{array}{l}
			\sum\limits_{k = 1}^K {w_k}\frac{{\left( {1 - \eta } \right)}}{{\ln 2}}\left[ {\ln \left( {1 + {\chi _k}} \right) - {\alpha _k}G\left( {{\chi _k}} \right)} \right]\\
			\ge \sum\limits_{k = 1}^K \left[ {\ln {{\left( {{\chi _k}} \right)}^{\hat w_k^{\left( i \right)}}} + {{\tilde w}_k}\delta _k^{\left( i \right)} - {{\tilde w}_k}{\alpha _k}\hat \delta _k^{\left( i \right)}} \right],
		\end{array}
	\end{equation}
	where ${{\hat w}^{\left( i \right)}_k}$ is  ${w_k}\frac{{\left( {1 - \eta } \right)}}{{\ln 2}}\left( {{\rho ^{\left( i \right)}} - {\alpha _k}{{\hat \rho }^{\left( i \right)}}} \right)$, and ${\tilde{w}_k}$ is ${w_k}\frac{{\left( {1 - \eta } \right)}}{{\ln 2}}$.}

Based on (\ref{downlink_WSR}), maximizing the LB in (\ref{downlink_WSR}) is equivalent to maximizing the function of $\ln {{\left( {{\chi _k}} \right)}^{{{\hat w}^{\left( i \right)}_k}}}$, as the constant term $({w_k}\frac{{\left( {1 - \eta } \right)}}{{\ln 2}}{\delta ^{\left( i \right)}_{k}} - {w_k}\frac{{\left( {1 - \eta } \right)}}{{\ln 2}}{\alpha _k}{{\hat \delta }^{\left( i \right)}_{k}})$ can be ignored. Furthermore, since $\ln {{\left( {{\chi _k}} \right)}^{{{\hat w}^{\left( i \right)}_k}}}$ is a monotonically increasing function of ${{\chi _k}}$, the maximal value of $\ln {{\left( {{\chi _k}} \right)}^{{{\hat w}^{\left( i \right)}_k}}}$ can be obtained by maximizing ${{\left( {{\chi _k}} \right)}^{{{\hat w}^{\left( i \right)}_k}}}$. As a result, we can maximize the following subproblem in the $i$th iteration instead of the original complex form, which is given by
\begin{subequations}
\setlength\abovedisplayskip{5pt}
\setlength\belowdisplayskip{5pt}
\label{equal_formualtion}
\begin{align}
	\mathop {\max }\limits_{\left\{ {p_k^p} \right\},\left\{ {p_{m,k}^d} \right\},\left\{ {{\chi _k}} \right\}} & \prod\limits_{k = 1}^K {{\chi _k}^{{{\hat w}_k^{\left( i \right)}}}}  \label{equal_formualtion_a}\\
	{\text{s.t.}}\;\;\;\;\;\;\;\; & {\rm{(\ref{downlink_obj_trans_b})}}, {\rm{(\ref{downlink_obj_trans_c})}}, {\rm{(\ref{downlink_obj_trans_d})}} \label{equal_formualtion_b}.
\end{align}
\end{subequations}
The above problem is not in the GP form, since the left hand side of constraint (\ref{downlink_obj_trans_b}) is a polynomial function. To address this, we rewrite the $k$th device's SINR expression, and then transform the constraint (\ref{downlink_obj_trans_b}) into a more tractable form.

\begin{lemma}
	\label{MRT_trans}
	The $k$th device's SINR can be rewritten as
	\begin{equation}
		\setlength\abovedisplayskip{5pt}
		\setlength\belowdisplayskip{5pt}
		\label{MRT_eq}
		\hat \gamma _k^{d} \notag =\frac{{N{{\left( {{\varphi _k}} \right)}^2}\prod\limits_{k' \in \left\{ {{{\cal Q}_k}\backslash k} \right\}} {{{\left( {{\theta _{k,k'}}} \right)}^2}} }}{{\rm{De}}_k},
	\end{equation}
	where ${\rm{De}}_k$ is given by
	\begin{align}
		\setlength\abovedisplayskip{5pt}
		\setlength\belowdisplayskip{5pt}
		&{\rm{De}}_k \notag \\
		& = \prod\limits_{k' \in \left\{ {{{\cal Q}_k}} \right\}} {{{\left( {{\theta _{k,k'}}} \right)}^2}} \left( {\sum\limits_{k' = 1}^K {\sum\limits_{m \in {{\cal M}_{k'}}} {p_{m,k'}^d{\beta _{m,k}}} } } \right)  \notag \\
		&\quad + N{{\left( {{\theta _{k,k}}} \right)}^2}\sum\limits_{k' \in \left\{ {{{\cal Q}_k}\backslash k} \right\}} {{{\left( {{\phi _{k,k'}}\prod\limits_{j \in \left\{ {{{\cal Q}_k}\backslash \left\{ {k,k'} \right\}} \right\}} {{\theta _{k,j}}} } \right)}^2}}  \label{De_k} \\
		&\quad + \prod\limits_{k' \in \left\{ {{{\cal Q}_k}} \right\}} {{{\left( {{\theta _{k,k'}}} \right)}^2}} \notag.
	\end{align}
	
	In (\ref{De_k}), ${{\varphi _k}}$, ${{\phi _{k,k'}}}$, and ${{\theta _{k,k'}}}$ are given by
	\begin{align}
		\setlength\abovedisplayskip{5pt}
		\setlength\belowdisplayskip{5pt}
		& {{\varphi _k}}\notag \\
		&\!\!=\!\!{\sum\limits_{m \in {{\cal M}_k}} \!\!\!{\sqrt {\tau p_k^pp_{m,k}^d{{\left( {{\beta _{m,k}}} \right)}^2}\!\!\!\!\!\!\!\!\prod\limits_{n \in \left\{ {{{\cal M}_k}\backslash m} \right\}}\!\!\!\!{\left( {\sum\limits_{i \in {{\cal Q}_k}} {\tau p_i^p{\beta _{n,i}}}  + 1} \right)} } } } 	\label{varphi_k},
	\end{align}
	\begin{align}
		\setlength\abovedisplayskip{5pt}
		\setlength\belowdisplayskip{5pt}
		&{{\phi _{k,k'}}}  \notag \\ 
		&\!\!=\!\!{\sum\limits_{m \in {{\cal M}_{k'}}}\!\!\!{\sqrt {\tau p_k^pp_{m,k'}^d{{\left( {{\beta _{m,k}}} \right)}^2}\!\!\!\!\!\!\!\!\prod\limits_{n \in \left\{ {{{\cal M}_{k'}}\backslash m} \right\}} \!\!\!\!{\left( {\sum\limits_{i \in {{\cal Q}_k}} {\tau p_i^p{\beta _{n,i}}}  + 1} \right)} } } } 	\label{phi_k},
	\end{align}
	and
	\begin{equation}
		\setlength\abovedisplayskip{5pt}
		\setlength\belowdisplayskip{5pt}
		\label{theta_k}
		{{\theta _{k,k'}}}= {\prod\limits_{m \in {{\cal M}_{k'}}} {\sqrt {\sum\limits_{i \in {{\cal Q}_k}} {\tau p_i^p{\beta _{m,i}}}  + 1} } }.
	\end{equation}
	
	{\emph{Proof}}: Please refer to Appendix \ref{MRT_trans_proof}. $\hfill\blacksquare$
\end{lemma}

Using Lemma \ref{MRT_trans}, the constraint (\ref{downlink_obj_trans_b}) can be equivalently transformed into the following inequality
\begin{equation}
	\setlength\abovedisplayskip{5pt}
	\setlength\belowdisplayskip{5pt}
	\label{downlink_inequality}
	{N{{\left( {{\varphi _k}} \right)}^2}\prod\limits_{k' \in \left\{ {{{\cal Q}_k}\backslash k} \right\}} {{{\left( {{\theta _{k,k'}}} \right)}^2}} } \ge \chi_k \times {\rm{De}}_k.
\end{equation}
However, the constraint (\ref{downlink_inequality}) does not satisfy the GP criterion, since both sides in (\ref{downlink_inequality}) are polynomial functions. To tackle this, we introduce the following theorem for approximating the numerator as a monomial function.

\begin{theorem}
	\label{theorem_MRT}
	For any given ${\hat p}_k^p >0$ and ${\hat p}_{m,k}^d > 0$, ${\varphi _k}\prod\limits_{k' \in \left\{ {{{\cal Q}_k}\backslash k} \right\}} {\left( {{\theta _{k,k'}}} \right)}$ is lower bounded by
	\begin{align}
		\setlength\abovedisplayskip{5pt}
		\setlength\belowdisplayskip{5pt}
		&{\varphi _k}\prod\limits_{k' \in \left\{ {{{\cal Q}_k}\backslash k} \right\}} {\left( {{\theta _{k,k'}}} \right)} \label{MRT_theorem} \\
		\ge& {c_k}\prod\limits_{m \in {{\cal M}_k}} {{{\left( {p_{m,k}^d{{\left( {{\beta _{m,k}}} \right)}^2}} \right)}^{{a_{m,k}}}}} \prod\limits_{i \in {{\cal Q}_k}} {{{\left( {\tau p_i^p} \right)}^{{b_i}}}} \notag,
	\end{align}
	where $b_i$ is given by (\ref{MRT_theorem_bi}) at the bottom of this page, $a_{m,k}$ and $c_k$ are given by
	\begin{figure*}[hb]
		\hrulefill
	\begin{equation}
		\setlength\abovedisplayskip{5pt}
		\setlength\belowdisplayskip{5pt}
		\label{MRT_theorem_bi}
		{b_i} = \left\{ {\begin{array}{*{20}{c}}
				\begin{array}{l}
					\frac{{\sum\limits_{m \in {{\cal M}_k}} {\frac{{{\hat p}_{m,k}^d{{\left( {{\beta _{m,k}}} \right)}^2}\tau {\hat p}_k^p\prod\limits_{n \in \left\{ {{{\cal M}_k}\backslash m} \right\}} {\tau {\hat p}_i^p{\beta _{n,i}}\sum\limits_{j \in \left\{ {{{\cal M}_k}\backslash m,n} \right\}} {\left( {\sum\limits_{i' \in {{\cal Q}_k}} {\tau {\hat p}_{i'}^p{\beta _{j,i'}}}  + 1} \right)} } }}{{\sqrt {\prod\limits_{n \in \left\{ {{{\cal M}_k}\backslash m} \right\}} {{\hat p}_{m,k}^d{{\left( {{\beta _{m,k}}} \right)}^2}\tau {\hat p}_k^p\left( {\sum\limits_{i' \in {{\cal Q}_k}} {\tau {\hat p}_{i'}^p{\beta _{n,i'}}}  + 1} \right)} } }}} }}{{2{{\hat \varphi }_k} }}\\
					+ \sum\limits_{k' \in \left\{ {{{\cal Q}_k}\backslash k} \right\}} {\sum\limits_{m \in {{\cal M}_{k'}}} {\frac{{\tau {\hat p}_{i}^p{\beta _{m,i}}}}{{2\left( {\sum\limits_{i' \in {{\cal Q}_k}} {\tau {\hat p}_{i'}^p{\beta _{m,i'}}}  + 1} \right)}}} }
				\end{array}&{i \ne k}\\
				\begin{array}{l}
					\frac{1}{2} + \frac{{\sum\limits_{m \in {{\cal M}_k}} {\frac{{{\hat p}_{m,k}^d{{\left( {{\beta _{m,k}}} \right)}^2}\tau {\hat p}_k^p\prod\limits_{n \in \left\{ {{{\cal M}_k}\backslash m} \right\}} {\tau {\hat p}_i^p{\beta _{n,i}}\sum\limits_{j \in \left\{ {{{\cal M}_k}\backslash m,n} \right\}} {\left( {\sum\limits_{i' \in {{\cal Q}_k}} {\tau {\hat p}_{i'}^p{\beta _{j,i'}}}  + 1} \right)} } }}{{\sqrt {\prod\limits_{n \in \left\{ {{{\cal M}_k}\backslash m} \right\}} {{\hat p}_{m,k}^d{{\left( {{\beta _{m,k}}} \right)}^2}\tau {\hat p}_k^p\left( {\sum\limits_{i' \in {{\cal Q}_k}} {\tau {\hat p}_{i'}^p{\beta _{n,i'}}}  + 1} \right)} } }}} }}{{2{{\hat \varphi }_k} }}\\
					+ \sum\limits_{k' \in \left\{ {{{\cal Q}_k}\backslash k} \right\}} {\sum\limits_{m \in {{\cal M}_{k'}}} {\frac{{\tau {\hat p}_{i}^p{\beta _{m,i}}}}{{2\left( {\sum\limits_{i' \in {{\cal Q}_k}} {\tau {\hat p}_{i'}^p{\beta _{m,i'}}}  + 1} \right)}}} }
				\end{array}&{i = k}
		\end{array}} \right.,
	\end{equation}
	\end{figure*}

	\begin{align}
		\setlength\abovedisplayskip{5pt}
		\setlength\belowdisplayskip{5pt}
		a_{m,k}= \frac{{\sqrt {\!\!\prod\limits_{n \in \left\{ {{{\cal M}_k}\backslash m} \right\}} \!\!\!\!{ {{\hat p}_{m,k}^d{{\left( {{\beta _{m,k}}} \right)}^2}}\tau {\hat p}^{p}_k\left( {\sum\limits_{i \in {{\cal Q}_k}} { \tau {\hat p}^{p}_i {\beta _{n,i}}}  + 1} \right)} } }}{{2{{{\hat \varphi }_k}}}} 	\label{MRT_theorem_amk},
	\end{align}
	and
	\begin{equation}
		\setlength\abovedisplayskip{5pt}
		\setlength\belowdisplayskip{5pt}
		\label{MRT_theorem_ek}
		c_k = \frac{{{{\hat \varphi }_k}\prod\limits_{k' \in \left\{ {{{\cal Q}_k}\backslash k} \right\}} {\left( {{{\hat \theta }_{k,k'}}} \right)} }}{{\prod\limits_{m \in {{\cal M}_k}} {{{\left( {\hat p_{m,k}^d{{\left( {{\beta _{m,k}}} \right)}^2}} \right)}^{{a_{m,k}}}}} \prod\limits_{i \in {{\cal Q}_k}} {{{\left( {\tau \hat p_i^p} \right)}^{{b_i}}}} }}.
	\end{equation}
	Furthermore, ${{\hat \varphi} _k}$ and ${{{\hat \theta} _{k,k'}}}$ are obtained by substituting $p_{m,k}^d = {\hat p}_{m,k}^d$ and $p_{k}^p = {\hat p}_{k}^p$ into (\ref{varphi_k}) and (\ref{phi_k}), respectively.
	
	{\emph{Proof}}: Please refer to Appendix \ref{Down_proof}. $\hfill\blacksquare$
\end{theorem}

Based on Theorem \ref{theorem_MRT}, the numerator of (\ref{MRT_eq}) can be lower bounded by the best monomial function, which is detailed as follows
\begin{align}
	\setlength\abovedisplayskip{5pt}
	\setlength\belowdisplayskip{5pt}
	&{\varphi _k}{\prod _{k' \in \left\{ {{Q_k}\backslash k} \right\}}}\left( {{\theta _{k,k'}}} \right) \notag \\
	\ge&c_k^{(n)}\prod\limits_{m \in {{\cal M}_k}} {{{\left( {p_{m,k}^d{{\left( {{\beta _{m,k}}} \right)}^2}} \right)}^{a_{m,k}^{(n)}}}} \prod\limits_{i \in {{\cal Q}_k}} {{{\left( {\tau p_i^p} \right)}^{b_i^{(n)}}}} \label{MRT_appro},
\end{align}
where ${{a_{m,k}^{(n)}}}$ and ${b_i^{(n)}}$ are obtained by using $p_{m,k}^d = {\hat p}_{m,k}^{d,(n)}$ and $p_{k}^p = {\hat p}_{k}^{p,(n)}$ in the $n$th iteration. Then, we substitute the LB of the numerator into constraint (\ref{downlink_obj_trans_b}), and the optimization problem can be reformulated as
\begin{subequations}
	\setlength\abovedisplayskip{5pt}
	\setlength\belowdisplayskip{5pt}
	\label{final_formualtion_MRT}
	\begin{align}
		\mathop {\max }\limits_{\left\{ {p_k^p} \right\},\left\{ {p_{m,k}^d} \right\},\left\{ {{\chi _k}} \right\}} & \prod\limits_{k = 1}^K {{\chi _k}^{{{\hat w}_k^{\left( i \right)}}}}  \label{final_MRT_a}\\
		{\text{s.t.}}\;\;\;\;\;\;\;\; & 	N ({c_k^{(n)}})^2 \prod\limits_{m \in {{\cal M}_k}} {{{( {p_{m,k}^d{{( {{\beta _{m,k}}} )}^2}} )}^{{2a_{m,k}^{(n)}}}}}  \notag \\  
		\times&\prod\limits_{i \in {{\cal Q}_k}} {{{\left( {\tau p_i^p} \right)}^{{2b_i^{(n)}}}}} \ge \chi_k \times {\rm{De}}_k, \label{final_MRT_b} \\ &{\rm{(\ref{downlink_obj_trans_c})}}, {\rm{(\ref{downlink_obj_trans_d})}} \label{final_MRT_c}.
	\end{align}
\end{subequations}
Based on the above-mentioned discussions, the power allocation can be transformed into a GP problem, which can be readily solved by CVX.

Since the WSR maximization is an NP-hard problem, we have to find a feasible region for initializing the algorithm. To tackle this, we construct an alternative optimization problem to find a feasible region for both the pilot power and payload power, which is given by
\begin{subequations}
	\setlength\abovedisplayskip{5pt}
	\setlength\belowdisplayskip{5pt}
	\label{feasible_region}
	\begin{align}
		\mathop {\max } \limits_{\rho ,\left\{ {p_{k}^p} \right\},\left\{ {p_{m,k}^d} \right\}}  & \rho \label{feasible_region_a} \\
		{\text{s.t.}}\;\;\;\;\;  & 		N ({c_k^{(n)}})^2 \prod\limits_{m \in {{\cal M}_k}} {{{( {p_{m,k}^d{{( {{\beta _{m,k}}} )}^2}} )}^{{2a_{m,k}^{(n)}}}}} \notag \\  
		 \times&\prod\limits_{i \in {{\cal Q}_k}} {{{\left( {\tau p_i^p} \right)}^{{2b_i^{(n)}}}}} \ge \frac{\rho }{{f_k^{ - 1}\left( {\frac{{R_k^{{\rm{req}}}\ln 2}}{{1 - \eta }}} \right)}} \times {\rm De}_k, \label{feasible_region_b} \\
		& ({\rm{\ref{downlink_obj_trans_d}}}) \label{feasible_region_c}.
	\end{align}
\end{subequations}
Problem (\ref{feasible_region}) is a GP problem and it is always feasible. Furthermore, the resource allocation can only be solved, when $\rho$ is no smaller than 1. Based on the abovementioned discussions, our iterative algorithm conceived for resource allocation is given in Algorithm \ref{MRT_algorithm}.


\begin{algorithm}[t]
	\caption{Iterative Algorithm for Maximizing WSR}
	\begin{algorithmic}[1]
		\label{MRT_algorithm}
		\STATE Initialize the iteration number $n = 1$, and error tolerance $\zeta = 0.01$;
		\STATE Based on the pilot allocation strategy $\mathfrak{L}_{\rm{pro}}$, initialize the pilot power $\left\{p_k^p = p_k^{p,(1)},\forall k\right\}$, calculate the transmission power $\left\{ p^{d,\left( 1\right)}_{m,k},\forall m, k \right\}$ by solving Problem (\ref{feasible_region}), obtain SINR $\left \{\chi _k^{\left( 1 \right)},\forall k\right\}$ and the WSR in (\ref{downlink_obj}) denoted as ${\rm{Obj}}^{\left(1\right)}$. Set ${\rm{Obj}}^{\left(0\right)} = {\rm{Obj}}^{\left(1\right)} \zeta$;
		\WHILE {${{\left( {{\rm{Ob}}{{\rm{j}}^{\left( n \right)}} - {\rm{Ob}}{{\rm{j}}^{\left( {n - 1} \right)}}} \right)} \mathord{\left/
					{\vphantom {{\left( {{\rm{Ob}}{{\rm{j}}^{\left( n \right)}} - {\rm{Ob}}{{\rm{j}}^{\left( {n - 1} \right)}}} \right)} {{\rm{Ob}}{{\rm{j}}^{\left( {n - 1} \right)}}}}} \right.
					\kern-\nulldelimiterspace} {{\rm{Ob}}{{\rm{j}}^{\left( {n - 1} \right)}}}} \ge \zeta$}
		\STATE Update $\left \{ {{\hat w}^{\left( n \right)}_k},c^{\left(n\right)}_k, a^{\left(n\right)}_{m,k}, b^{(n)}_k,\forall m,k \right\}$;
		\STATE Update $n = n+1$, solve Problem (\ref{final_formualtion_MRT}) by using the CVX package to obtain $\left \{p_k^{p,(n)},p_{m,k}^{d,\left( n \right)} ,\forall m,k\right\}$, calculate SINR $\left \{\chi _k^{\left( n \right)},\forall k\right\}$ and then obtain the WSR, denoted as ${\rm{Obj}}^{\left(n\right)}$;
		\ENDWHILE
	\end{algorithmic}
\end{algorithm}

\subsection{Algorithm analysis}
The convergence of Algorithm \ref{MRT_algorithm} can be readily proved by using a similar process to that in \cite{ref13d}, and thus it is omitted here.  \textcolor{black}{Then, we analyze the complexity of our proposed algorithm. Specifically, the main complexity of each iteration in Algorithm 2 lies in solving Problem (26) which includes $(2+M)K$ variables and $2K + M$ constraints. Based on \cite{van2018joint}, the computational complexity of this algorithm is on the order of ${\mathcal{O}}(N_{iter} \times \max\{[(M + 2)K]^{3},(2K + M)[(M + 2)K]^{2}, N_{cost}\})$, where $N_{iter}$ is the number of iterations and $N_{cost}$ is the computational complexity of calculating the first-order and second-order derivatives of the objective function and constraint functions of Problem (26).}

\section{Simulation Results}
In this section, the performance of our proposed pilot allocation strategy and power control algorithm is numerically evaluated and discussed.

\subsection{Simulation Parameters}
There are $M$ APs that are uniformly positioned constellation points in a smart factory of size $0.2 \times 0.2$ ${\text{km}}^2$ and $K$ devices that are independently and uniformly distributed. The large-scale fading factors are based on the Hata-COST231 propagation model of \cite{tang2001mobile}, where the heights of APs and the devices are 15 m and 1.6 m, respectively. Specifically, the three-slope path loss (dB) can be expressed as
\begin{equation}
	\setlength\abovedisplayskip{5pt}
	\setlength\belowdisplayskip{5pt}
	\label{channel_model}
	{{\rm{PL}}_{m,k}} \!\!= \!\! \left\{ {\begin{array}{*{20}{l}}
			 \!\! \!\!{{L_{\rm ls}}\! +\! 35{\log_{10}}({d_{m,k}}),}&\!\! \!\!{{d_1} < {d_{m,k}},}\\
			 \!\! \!\!{{L_{\rm ls}} \!+\! 15{\log_{10}}({d_1})\! +\! 20{\log_{10}}({d_{m,k}}),}& \!\!\!\!{{d_0} < {d_{m,k}} \le {d_1},}\\
			 \!\! \!\!{{L_{\rm ls}} \!+\! 15{\log_{10}}({d_1}) \!+\! 20{\log_{10}}({d_0}),}& \!\!\!\!{{d_{m,k}} \le {d_0},} \\
			 
	\end{array}} \right.
\end{equation}
where $d_{m,k}$ is the distance between the $m$th AP and the $k$th device, $L_{\rm{ls}}$ is a constant of 140.7 (dB), while $d_0$ and $d_1$ are 0.01 km and 0.05 km, respectively. \textcolor{black}{As for small-scale fading, it is generally modeled as Rayleigh fading with zero mean and unit variance. These parameter are similar to those in \cite{ref16,ref14,2019ce}}. 

The corresponding normalized pilot power $p_k^p$ and payload power $p_{m,k}^d$ can be computed through dividing these powers by the noise power, where the noise power is given by
\begin{equation}
	\setlength\abovedisplayskip{5pt}
	\setlength\belowdisplayskip{5pt}
	\label{noise_power}
	P_n = B \times {k_B} \times {T_0} \times  10^{\frac{9}{10}} \left( {\rm{W}} \right),
\end{equation}
where $k_B = 1.381 \times 10^{-23}$ (Joule per Kelvin) is the Boltzmann constant, and ${T_0} = 290$ (Kelvin) is the noise temperature. We set the bandwidth to $1$ MHz and the transmission latency to 100 us. Thus, the total transmission blocklength is $L = 100$, where the blocklength fraction of $\tau$ is utilized for channel estimation and the remaining $(L - \tau)$ blocklength fraction is used for downlink payload transmission \cite{ozdogan2019performance}. The devices' weights are randomly generated within [0,1]. Moreover, unless otherwise specified, the DEP requirements of all devices and the data rates are $10^{-7}$ and $0.75$ $\text{bit/s/Hz}$, respectively.

Again, the user-centric approach can reduce the implementation complexity. Similar to \cite{ref14}, the AP selection based on the pathloss/distance is adopted. More specifically, the pathloss factors from the $k$th device to all APs $\{\beta_{1,k},\beta_{2,k}, \cdot \cdot \cdot, \beta_{M,k} \}$ are sorted in a descending order, and then we select the specific factors until satisfying the following condition
\begin{equation}
	\setlength\abovedisplayskip{5pt}
	\setlength\belowdisplayskip{5pt}
	\label{AP_selection}
	\frac{{\sum\limits_{m \in {{\cal M}_k}} {{\beta _{m,k}}} }}{{\sum\limits_{m = 1}^M {{\beta _{m,k}}} }} \ge {T_h},
\end{equation}
where $T_h$ is the threshold. Then, the set of ${\cal{U}}_m$ is obtained by checking whether the $m$th AP belongs to the set of ${\cal{M}}_k$, $\forall k \in \{ 1,2, \cdot \cdot \cdot, K\}$. A higher threshold indicates that a device is served by more APs, which enhances the performance at the cost of increased complexity. Unless otherwise stated, $T_h$ is set to 0.75 for all the following simulations for striking a balance between the performance vs. implementation complexity trade-off.

\subsection{The LB data rate and ergodic data rate}
\begin{figure*}[ht]
	\setlength{\abovecaptionskip}{-5pt}
	\setlength{\belowcaptionskip}{-15pt}
	\centering
	\begin{minipage}[t]{0.45\linewidth}
			\centering
			\includegraphics[width= 1\textwidth]{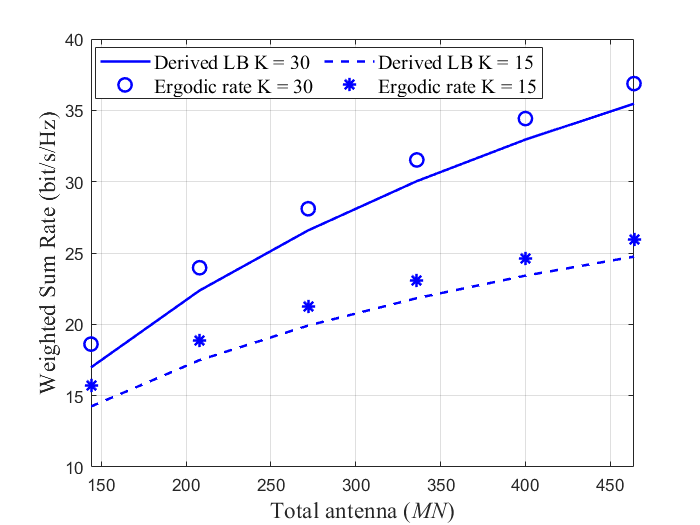}
			\DeclareGraphicsExtensions.
			\captionsetup{font={small}}
			\caption{WSR in Downlink system V.S. The Number of Total Antennas $MN$ with $p_k^p$ = 0.1 W, $p_{m,k}^d = \frac{0.2}{|{\cal{U}}_m|}$ W, $\forall m$ and $\forall k \in {\mathcal {U}_m}$.}
			\label{Downlink_LB}
		\end{minipage}
	\hspace{5mm}
	\begin{minipage}[t]{0.45\linewidth}
		\centering
		\includegraphics[width= 1\textwidth]{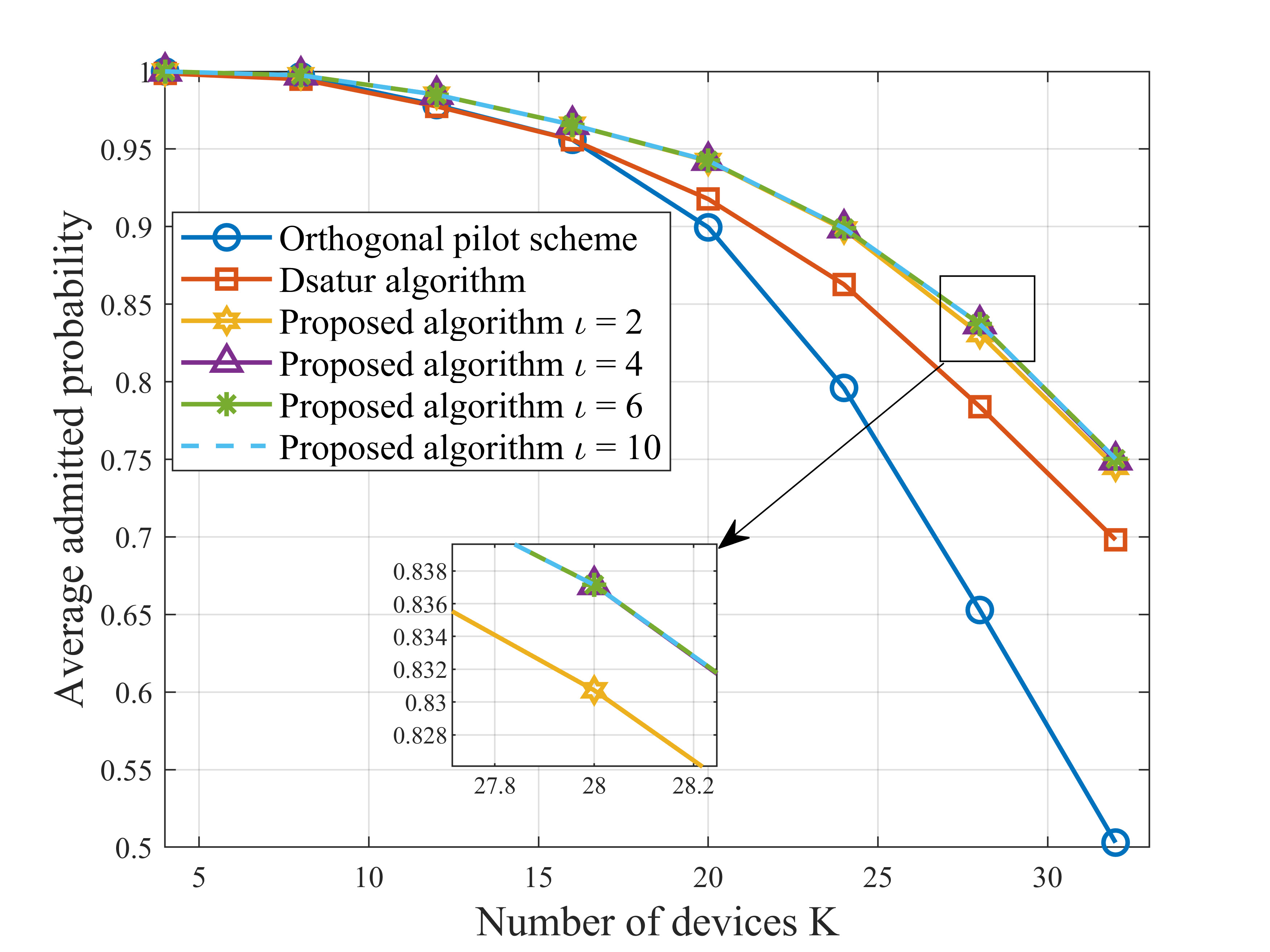}
		\DeclareGraphicsExtensions.
		\captionsetup{font={small}}
		\caption{Average available probability of devices with $M = 16$, $N = 9$, $n_{\rm{max}} = 4$, $N_{\max}^{\rm iter} = 20$, $P_k^{p,\rm max} = 0.1$ W, and $p_{m,k}^d = \frac{0.2}{|{\cal U}_m|}$ W, $\forall k \in {\cal U}_m$.}
		\label{avilable_pilot}
	\end{minipage}
\end{figure*}
To check whether the LB can provide a more convenient expression for resource allocation, in Fig. \ref{Downlink_LB}, we first evaluate the gap between the LB data rate and the achievable ergodic downlink data. The results are obtained through $10^4$ Monte-Carlo simulations with $M$ = 16. The WSRs based on the rate derived can approach the achievable erogdic data rate for any given $N$ and $K$, which demonstrates the efficiency of our resource allocation based on the expressions derived.

\subsection{Admitted devices based on proposed pilot assignment}
\begin{figure*}[ht]
	\setlength{\abovecaptionskip}{-5pt}
	\setlength{\belowcaptionskip}{-15pt}
	\centering
	\begin{minipage}[t]{0.45\linewidth}
		\centering
		\includegraphics[width= 1\textwidth]{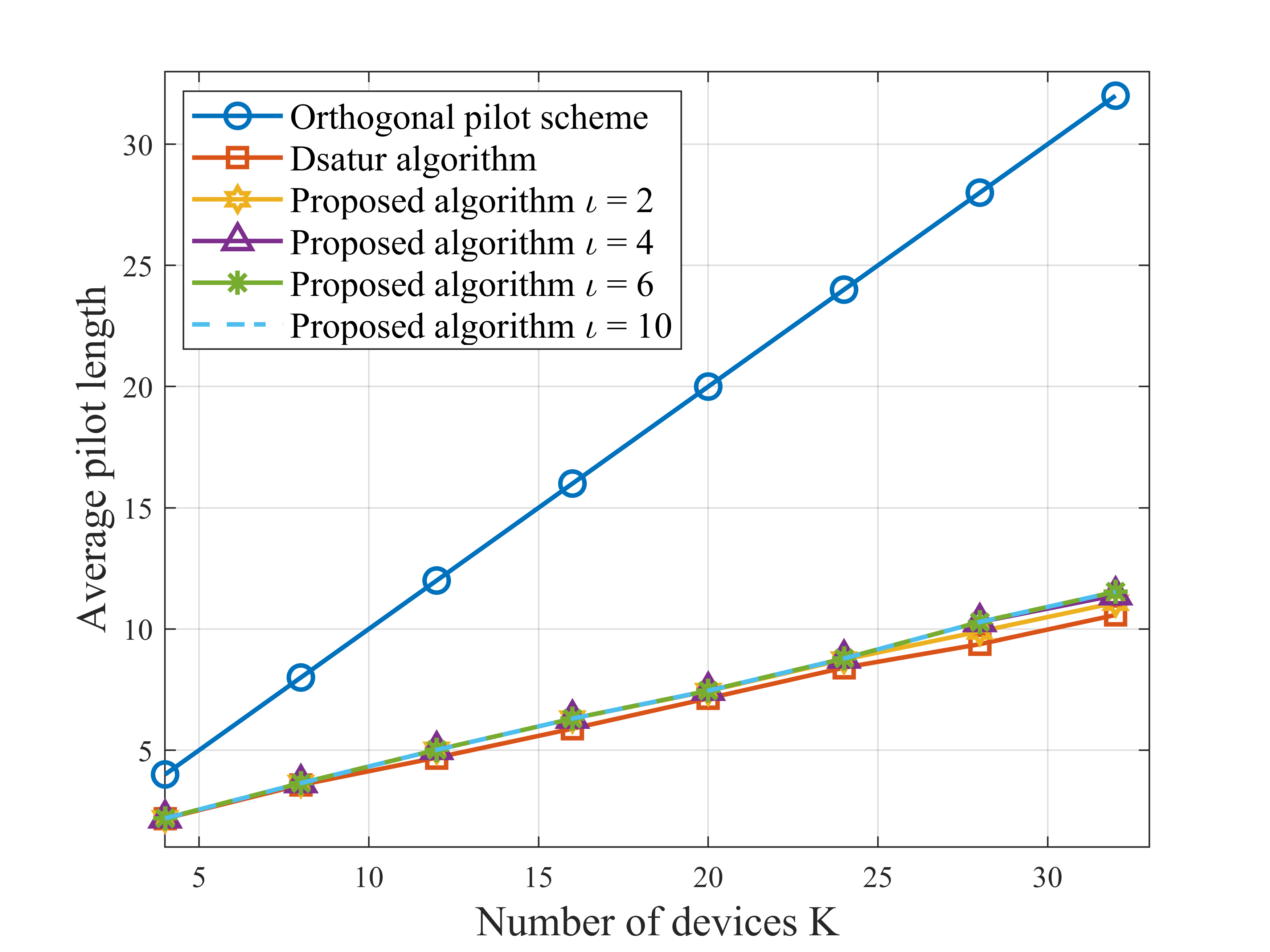}
		\DeclareGraphicsExtensions.
		\captionsetup{font={small}}
		\caption{Average pilot length for various pilot allocation strategies with $M = 16$, $N = 9$, $n_{\rm{max}} = 4$, $N_{\max}^{\rm iter} = 20$, $P_k^{p,\rm max} = 0.1$ W, and $p_{m,k}^d = \frac{0.2}{|{\cal U}_m|}$ W, $\forall k \in {\cal U}_m$.}
		\label{length_pilot}
	\end{minipage}
	\hspace{5mm}
		\begin{minipage}[t]{0.45\linewidth}
		\centering
		\includegraphics[width= 1\textwidth]{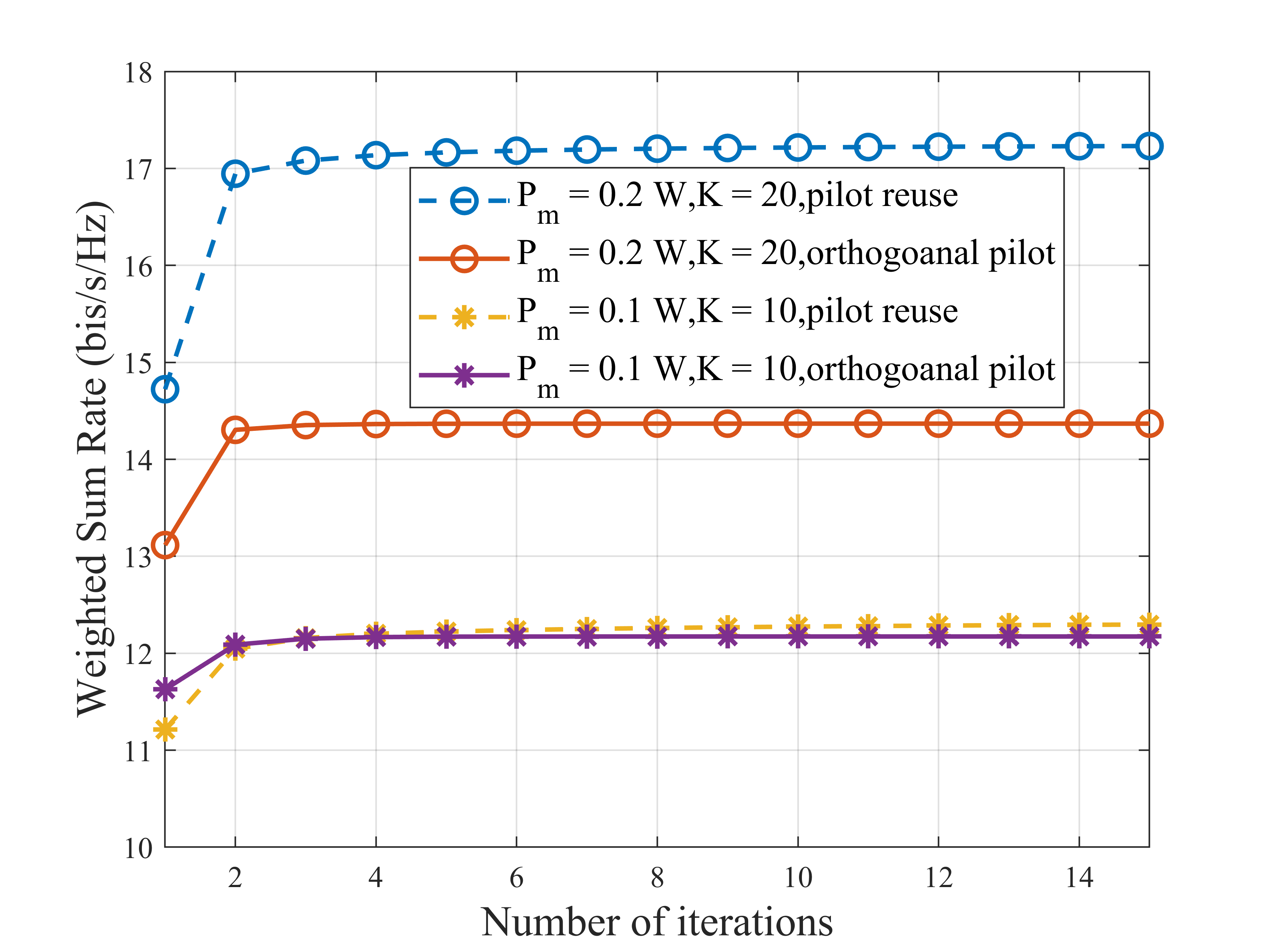}
		\DeclareGraphicsExtensions.
		\captionsetup{font={small}}
		\caption{Convergence of proposed algorithm for downlink system with $M = 16$, $N = 9$, and $P_k^{\max,p}$ = 0.1 W, $\forall k$.}
		\label{Downlink_Convergence}
	\end{minipage}
\end{figure*}
We fix the pilot power and transmission power to investigate how the pilot assignment strategy influences the devices' rate and DEP. By averaging the results over 1000 randomly generated locations, Fig. \ref{avilable_pilot} and Fig. \ref{length_pilot} depict the admitted device probability $\frac{|S|}{K}$ and the average pilot length $\tau$. It is observed that the admitted device probability decreases upon increasing the number of devices due to the increased interference. For the orthogonal scheme (i.e. each device is assigned a unique pilot sequence that is orthogonal to the other devices) in Fig. \ref{avilable_pilot}, we observe a significant drop trend when the number of devices $K$ exceeds 20, since some devices fail to satisfy the data rate requirements due to reducing the payload blocklength used for data transmission. By contrast, upon reducing the pilot overhead, an increased blocklength can be utilized for payload transmission and the number of admitted devices is enhanced accordingly. However, this implies that we cannot guarantee the devices' rate and DEP requirements by providing efficiently accuracy estimated channels, especially for large $K$. More importantly, our proposed approach is capable of significantly increasing the number of admitted devices over that of the Dsatur algorithm via optimizing the pilot fraction ($\tau$) and improving pilot allocation strategy. Furthermore, the optimal search region is $\iota = 4$, and thus we set $\iota = 4$ for all the following simulations to strike a performance vs. implementation complexity trade-off.

\subsection{Convergence of proposed algorithm}
%

Based on the given pilot strategy, we evaluate the convergence of Algorithm \ref{MRT_algorithm} for the power aided downlink systems, as illustrated in Fig. \ref{Downlink_Convergence}. We observe that it takes only 3 or 4 iterations for our algorithm to converge to a locally optimal solution, which verifies the efficiency of our approach.

\subsection{Performance of proposed algorithm}
\begin{figure*}[ht]
	\setlength{\abovecaptionskip}{-5pt}
	\setlength{\belowcaptionskip}{-5pt}
	\centering
	\begin{minipage}[t]{0.45\linewidth}
		\centering
		\includegraphics[width= 1\textwidth]{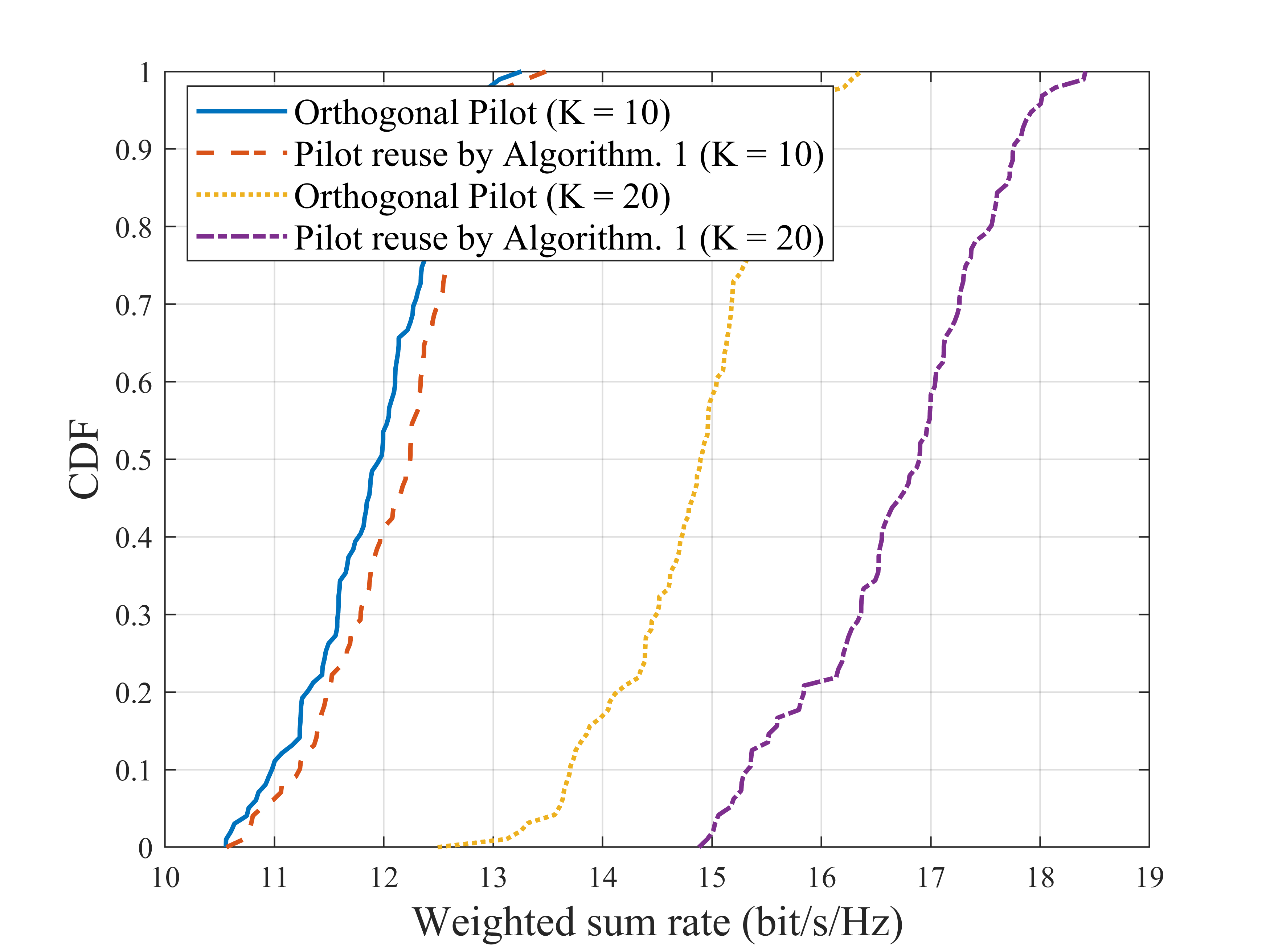}
		\DeclareGraphicsExtensions.
		\captionsetup{font={small}}
		\caption{\textcolor{black}{Cumulative distribution of WSR for downlink system with $M =$ 16, $N =$ 9, $P_m = $ 0.2 W, $\forall m$, $P_k^{{\rm max},p} = $ 0.1 W, and $R_k^{\rm req} = $ 0.5 bit/s/Hz, $\forall k$.}}
		\label{CDF_downlink}
	\end{minipage}
	\hspace{5mm}
		\begin{minipage}[t]{0.45\linewidth}
		\centering
		\includegraphics[width= 1\textwidth]{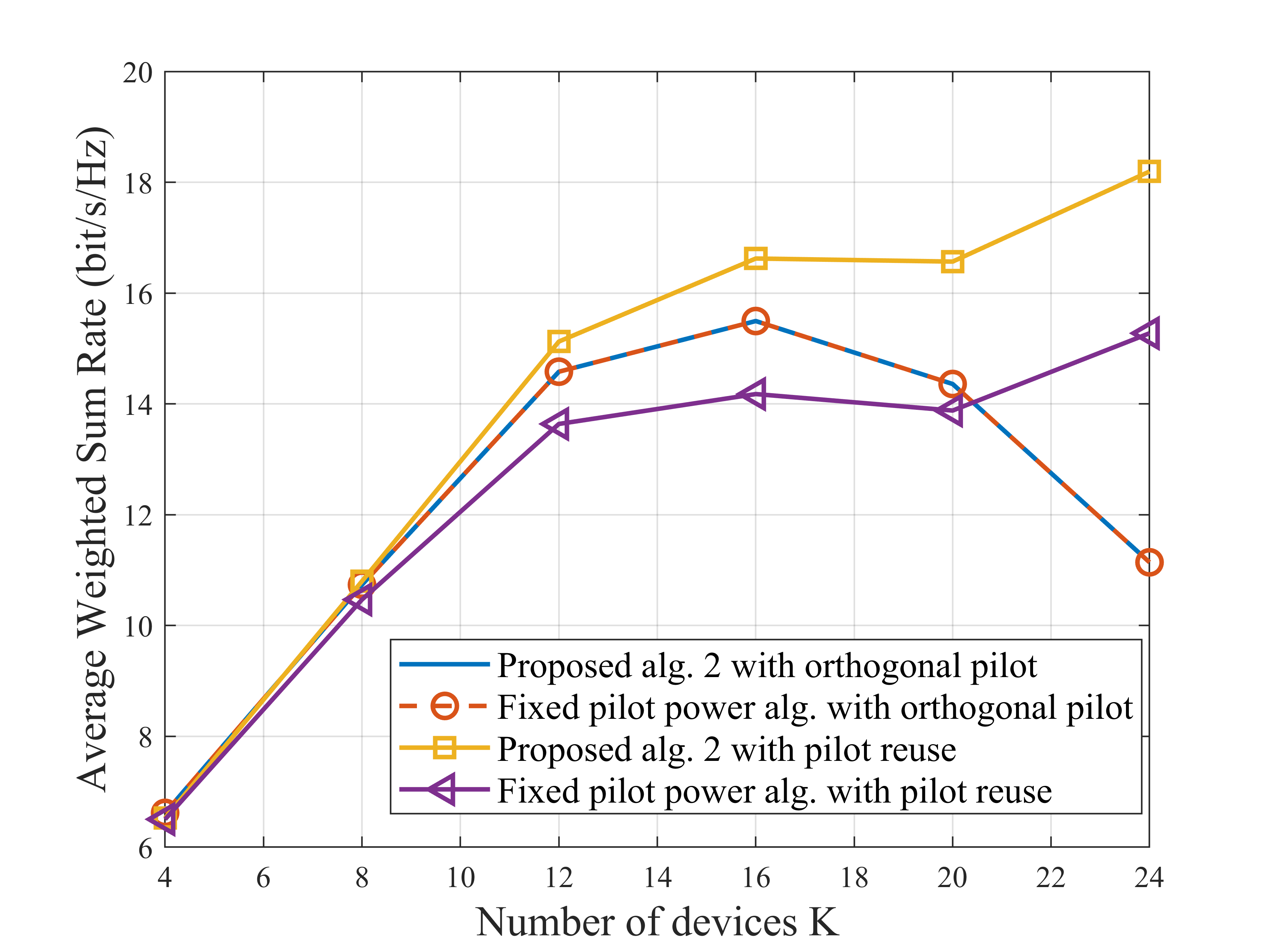}
		\DeclareGraphicsExtensions.
		\captionsetup{font={small}}
		\caption{Average WSR V.S. Number of Devices for downlink system  with $M = 16$, $N = 9$, $P_m = $ 0.2 W, $\forall m$, $P_k^{\rm max} = $ 0.1 W, and $R_k^{\rm req} = $ 0.5 bit/s/Hz, $\forall k$.}
		\label{AWSR_downlink}
	\end{minipage}
\end{figure*}
In Fig. \ref{CDF_downlink}, we investigate the performance of Algorithm \ref{Pilot_Assignment} and Algorithm \ref{MRT_algorithm}, by depicting the cumulative distribution function (CDF) of WSR for the  downlink throughput. Besides, if there is no feasible region for Algorithm \ref{MRT_algorithm} (e.g., some devices cannot satisfy the DEP and rate requirements), these results will not be listed. When $K = $ 10, sharing the pilot sequence in the CF mMIMO system slightly improves the WSR. By contrast, the considerable improvement on the system performance is obtained as the number of devices increases (i.e. $K$ = 20). This suggests that multiplexing the pilots is not essential when blocklength is adequate. Conversely, with the increasing number of devices, the blockelength is insufficient to satisfy the devices' requirements, and thus it is more efficient to set aside an increased payload blocklength for data transmission through sharing pilot sequences.

\subsection{Effect of number of Devices}

To further explore the impact of the number of devices on the system performance, Fig. \ref{AWSR_downlink} shows the average WSR of the downlink system by averaging 100 simulations. To guarantee all devices' rate and DEP requirements, if any device fails to satisfy the minimal requirements, the system performance is set to zero. Besides, we compare the average WSR to that of the fixed pilot power algorithm, where the pilot power is fixed to $p_k^p = P_k^{p,{\rm max}}, \forall k $.

It is observed from Fig. \ref{AWSR_downlink} for the CF mMIMO based orthogonal pilot-aided scheme, that the WSR increases up to $K = 16$, but beyond that it decays as $K$ increases. This is due to the fact that the available blocklength $(L - \tau)$ decreases upon increasing the number of devices $K$, hence resulting in the violation of the devices' rate requirements. When the orthogonal pilot scheme is employed, it is interesting to observe that the performance with the fixed pilot algorithm is similar to that of the joint optimization, since the SINR is a monotonically increasing function of the pilot power. This inspires us to adopt the maximal pilot power for enhancing the system performance of the orthogonal pilot scheme. By contrast, using the maximal pilot power for the system relying on pilot reuse fails to improve the performance, since the pilot contamination is increased with the pilot power. More importantly, it is observed that systems using orthogonal pilot sequences can perform better for a low system load (i.e. $K < 6$), while, for a high $K$, it is more effective to share the pilot sequences as well as to carefully allocate the pilot vs. payload fraction. 


\section{Conclusions}
The resource allocation problem of a CF mMIMO-aided URLLC system was investigated. We first derived the closed-form LB data rates with imperfect CSI and pilot contamination. Then, to guarantee the devices' rate and DEP, a novel pilot allocation scheme was proposed to balance the pilot vs. payload fraction of the blocklength with the aim of admitting more devices. Finally, by jointly optimizing the pilot and payload power to strike a trade-off between the estimated channel gain and pilot contamination, we transformed the non-convex problem into a series of subproblems, which can be solved in an iterative manner by the proposed algorithm. Our simulation results demonstrated the significant improvement in the number of admitted devices and the rapid convergence of the proposed algorithm. Essentially, CF mMIMO employing pilot reuse can support more devices than that with orthogonal pilots.


\begin{appendices}
\section{Proof of Theorem \ref{LB_MRT_T}}
\label{Proof_MRT_LB}

Before proving this theorem, we have to calculate the precoding vector for the MRT scheme. The normalized precoding vector is given by
\begin{equation}
	\setlength\abovedisplayskip{5pt}
	\setlength\belowdisplayskip{5pt}
	\label{MRT_normalized}
	\begin{split}
		&{\bf{a}}_{m,k}  \\
		& = \frac{{{\alpha _{m,k}}\left( {\sum\limits_{i \in {{\cal Q}_k}} {\sqrt {\frac{{p_i^p}}{{p_k^p}}} } {{\bf{g}}_{m,i}} + \frac{1}{{\sqrt {\tau p_k^p} }}{\bf{N}}_m^p{{\bf{q}}_k}} \right)}}{{\sqrt {{{ \mathbb{E} \left\| {{\alpha _{m,k}}\left( {\sum\limits_{i \in {{\cal Q}_k}} {\sqrt {\frac{{p_i^p}}{{p_k^p}}} } {{\bf{g}}_{m,i}} + \frac{1}{{\sqrt {\tau p_k^p} }}{\bf{N}}_m^p{{\bf{q}}_k}} \right)} \right\|}^2}} }} \\
		& = \frac{{\sum\limits_{i \in {{\cal Q}_k}} {\sqrt {\frac{{p_i^p}}{{p_k^p}}} } {{\bf{g}}_{m,i}} + \frac{1}{{\sqrt {\tau p_k^p} }}{\bf{N}}_m^p{{\bf{q}}_k}}}{{\sqrt {N\left( {\sum\limits_{i \in {{\cal Q}_k}} {\frac{{p_i^p}}{{p_k^p}}} {\beta _{m,i}} + \frac{1}{{\tau p_k^p}}} \right)} }} \\
		& = \frac{{\sqrt {{\lambda _{m,k}}} }}{{{\beta _{m,k}}\sqrt N }}\left( {\sum\limits_{i \in {{\cal Q}_k}} {\sqrt {\frac{{p_i^p}}{{p_k^p}}} } {{\bf{g}}_{m,i}} + \frac{1}{{\sqrt {\tau p_k^p} }}{\bf{N}}_m^p{{\bf{q}}_k}} \right),
	\end{split}
\end{equation}
where $\alpha _{m,k}$ is $\alpha _{m,k} = \frac{{\tau p_k^p{\beta _{m,k}}}}{{\sum\limits_{i \in {{\cal Q}_k}} {\tau p_i^p{\beta _{m,i}}}  + 1}}$.

Then, we derive the expressions of ${\left| {{\rm{DS}}_{k}^d} \right|^2}$, $\mathbb{E}\left( \left| {{\rm{LS}}_{k}^d} \right|^2 \right)$, $\mathbb{E}\left( \left| {{\rm{UI}}_{k,k'}^d} \right|^2 \right)$ and $\mathbb{E}\left( \left| {{\rm{N}}_{k}^d} \right|^2 \right)$, respectively. Since $\hat {\bf{g}}_{m,k}$ and $\tilde {\bf{g}}_{m,k}$ are independent, we have
\begin{equation}
	\setlength\abovedisplayskip{5pt}
	\setlength\belowdisplayskip{5pt}
	\label{MRC_DSk}
	\begin{split}
		 &{\left| {{\rm{DS}}_{k}^d} \right|^2}\\
		 & = {\left( \mathbb{E}{\left\{ {\sum\limits_{m \in {{\cal M}_k}} {{{\left( {{{\bf{g}}_{m,k}}} \right)}^T}{{({\bf{a}}_{m,k})}^ * }\sqrt {p_{m,k}^d} } } \right\}} \right)^2} \\
		& = {\left( \mathbb{E} {\left\{ {\sum\limits_{m \in {{\cal M}_k}} \frac{{\sqrt {{\lambda _{m,k}}} }}{{{\beta _{m,k}}\sqrt N }}{{{\left( {{{\bf{g}}_{m,k}}} \right)}^T}{{\left( {{{\bf{g}}_{m,k}}} \right)}^ * }\sqrt {p_{m,k}^d} } } \right\}} \right)^2} \\
		& = {\left( {\sum\limits_{m \in {{\cal M}_k}} {\sqrt {Np_{m,k}^d{\lambda _{m,k}}} } } \right)^2}.
	\end{split}
\end{equation}

Then, $\mathbb{E}\left( \left| {{\rm{UI}}_{k,k'}} \right|^2 \right)$ can be given by (\ref{MRC_UIkk}) at bottom of this page.
\begin{figure*}[hb]
	\hrulefill
\begin{equation}
	\setlength\abovedisplayskip{5pt}
	\setlength\belowdisplayskip{5pt}
	\label{MRC_UIkk}
	\begin{split}
		&\mathbb{E} \left( {{{\left| {{\rm{U}}{{\rm{I}}_{k,k'}}} \right|}^2}} \right)\\
		&=  \mathbb{E}\left\{ {{{\left| {\sum\limits_{m \in {{\cal M}_{k'}}} {\sqrt {p_{m,k'}^d} {{\left( {{{\bf{g}}_{m,k}}} \right)}^T}{{({\bf{a}}_{m,k'})}^ * }} } \right|}^2}} \right\}  \\
		& = \mathbb{E} \left\{{{{\left| {\sum\limits_{m \in {{\cal M}_{k'}}} {\frac{{\sqrt {p_{m,k'}^d{\lambda _{m,k'}}} }}{{{\beta _{m,k'}}\sqrt N }}\sqrt {\frac{{p_k^p}}{{p_{k'}^p}}} } {{\left( {{{\bf{g}}_{m,k}}} \right)}^T}{{\left( {{{\bf{g}}_{m,k}}} \right)}^ * }} \right|}^2}} \right\}  \\
		& \quad + \mathbb{E} \left\{{{{{{\left| {\sum\limits_{m \in {{\cal M}_{k'}}} {\sum\limits_{i \in \left\{ {{{\cal Q}_{k'}}\backslash k} \right\}} {\frac{{\sqrt {p_{m,k'}^d{\lambda _{m,k'}}} }}{{{\beta _{m,k'}}\sqrt N }}\sqrt {\frac{{p_i^p}}{{p_{k'}^p}}} {{\left( {{{\bf{g}}_{m,k}}} \right)}^T}{{\left( {{{\bf{g}}_{m,i}}} \right)}^ * }} } } \right|}^2}}}} \right\}  \\
		&\quad + \mathbb{E} \left\{{{{\left| {\sum\limits_{m \in {{\cal M}_{k'}}} {\frac{{\sqrt {p_{m,k'}^d{\lambda _{m,k'}}} }}{{{\beta _{m,k'}}\sqrt N }}\frac{1}{{\sqrt {\tau p_{k'}^p} }}{{\left( {{{\bf{g}}_{m,k}}} \right)}^T}{{\left( {{\bf{N}}_m^p{{\bf{q}}_{k'}}} \right)}^ * }} } \right|}^2}}\right\} .
	\end{split}
\end{equation}
\end{figure*}
For each term in (\ref{MRC_UIkk}), the first term and the second term are given by (\ref{UIkk_first}) and (\ref{UIkk_second}) at the bottom of this page and next page, respectively. Furthermore, the third term is denoted as 
\begin{figure*}[hb]
	\hrulefill
	\begin{equation}
		\setlength\abovedisplayskip{5pt}
		\setlength\belowdisplayskip{5pt}
		\label{UIkk_first}
		\begin{split}
			&\mathbb{E} \left\{ {{{\left| {\sum\limits_{m \in {{\cal M}_{k'}}} {\frac{{\sqrt {p_{m,k'}^d{\lambda _{m,k'}}} }}{{{\beta _{m,k'}}\sqrt N }}\sqrt {\frac{{p_k^p}}{{p_{k'}^p}}} } {{\left( {{{\bf{g}}_{m,k}}} \right)}^T}{{\left( {{{\bf{g}}_{m,k}}} \right)}^ * }} \right|}^2}} \right\} \\
			&= \frac{{p_k^p}}{{p_{k'}^p}}\mathbb{E}\left\{\sum\limits_{m \in {{\cal M}_{k'}}} {\frac{{p_{m,k'}^d{\lambda _{m,k'}}}}{{{{\left( {{\beta _{m,k'}}} \right)}^2}N}}{{\left( {{{\bf{g}}_{m,k}}} \right)}^T}{{\left( {{{\bf{g}}_{m,k}}} \right)}^ * }{{\left( {{{\bf{g}}_{m,k}}} \right)}^T}{{\left( {{{\bf{g}}_{m,k}}} \right)}^ * }}  \right\}\\
			& \quad+ \frac{{p_k^p}}{{p_{k'}^p}}\mathbb{E}\left\{\sum\limits_{m \in {{\cal M}_{k'}}} {\sum\limits_{m' \in \left\{ {{{\cal M}_{k'}}\backslash m} \right\}} {\frac{{\sqrt {p_{m,k'}^d{\lambda _{m,k'}}} }}{{{\beta _{m,k'}}\sqrt N }}\frac{{\sqrt {p_{m',k'}^d{\lambda _{m,k'}}} }}{{{\beta _{m',k'}}\sqrt N }}{{\left( {{{\bf{g}}_{m,k}}} \right)}^T}{{\left( {{{\bf{g}}_{m,k}}} \right)}^ * }{{\left( {{{\bf{g}}_{m',k}}} \right)}^T}{{\left( {{{\bf{g}}_{m',k}}} \right)}^ * }} }\right\} \\
			& = \frac{{p_k^p}}{{p_{k'}^p}}\left[ {\sum\limits_{m \in {{\cal M}_{k'}}} {\frac{{p_{m,k'}^d{\lambda _{m,k'}}{{\left( {{\beta _{m,k}}} \right)}^2}}}{{{{\left( {{\beta _{m,k'}}} \right)}^2}}}}  + N{{\left( {\sum\limits_{m \in {{\cal M}_{k'}}} {\sqrt {p_{m,k'}^d{\lambda _{m,k'}}} \frac{{{\beta _{m,k}}}}{{{\beta _{m,k'}}}}} } \right)}^2}} \right],
		\end{split}
	\end{equation}
\end{figure*}
\begin{figure*}[b]
	\hrulefill
	\begin{equation}
		\setlength\abovedisplayskip{5pt}
		\setlength\belowdisplayskip{5pt}
		\label{UIkk_second}
		\begin{split}
			& \mathbb{E} \left\{ {{{\left| {\sum\limits_{m \in {{\cal M}_{k'}}} {\sum\limits_{i \in \left\{ {{{\cal Q}_{k'}}\backslash k} \right\}} {\frac{{\sqrt {p_{m,k'}^d{\lambda _{m,k'}}} }}{{{\beta _{m,k'}}\sqrt N }}\sqrt {\frac{{p_i^p}}{{p_{k'}^p}}} {{\left( {{{\bf{g}}_{m,k}}} \right)}^T}{{\left( {{{\bf{g}}_{m,i}}} \right)}^ * }} } } \right|}^2}} \right\} \\
			& = \mathbb{E} \left\{ {\sum\limits_{m \in {{\cal M}_{k'}}} {\sum\limits_{i \in \left\{ {{{\cal Q}_{k'}}\backslash k} \right\}} {\frac{{p_{m,k'}^d{\lambda _{m,k'}}}}{{{{\left( {{\beta _{m,k'}}} \right)}^2}N}}\frac{{p_i^p}}{{p_{k'}^p}}{{\left| {{{\left( {{{\bf{g}}_{m,k}}} \right)}^T}{{\left( {{{\bf{g}}_{m,i}}} \right)}^ * }} \right|}^2}N{\beta _{m,k}}{\beta _{m,i}}} } } \right\} \\
			& = \sum\limits_{m \in {{\cal M}_{k'}}} {\frac{{p_{m,k'}^d{\lambda _{m,k'}}{\beta _{m,k}}}}{{{{\left( {{\beta _{m,k'}}} \right)}^2}}}\sum\limits_{i \in \left\{ {{{\cal Q}_{k'}}\backslash k} \right\}} {\frac{{p_i^p}}{{p_{k'}^p}}{\beta _{m,i}}} } ,
		\end{split}
	\end{equation}
\end{figure*}
\begin{equation}
	\setlength\abovedisplayskip{5pt}
	\setlength\belowdisplayskip{5pt}
	\label{UIkk_thrid}
	\begin{array}{l}
		\mathbb{E}\left\{ {{{\left| {\sum\limits_{m \in {{\cal M}_{k'}}} {\frac{{\sqrt {p_{m,k'}^d{\lambda _{m,k'}}} }}{{{\beta _{m,k'}}\sqrt N }}\frac{1}{{\sqrt {\tau p_{k'}^p} }}{{\left( {{{\bf{g}}_{m,k}}} \right)}^T}{{\left( {{\bf{N}}_m^p{{\bf{q}}_{k'}}} \right)}^*}} } \right|}^2}} \right\}\\
		= \frac{1}{{\tau p_{k'}^p}}\sum\limits_{m \in {{\cal M}_{k'}}} {\frac{{p_{m,k'}^d{\lambda _{m,k'}}{\beta _{m,k}}}}{{{{\left( {{\beta _{m,k'}}} \right)}^2}}}} .
	\end{array}
\end{equation}

Upon substituting (\ref{UIkk_first}), (\ref{UIkk_second}), and (\ref{UIkk_thrid}) into (\ref{MRC_UIkk}), we have
\begin{align}
	\setlength\abovedisplayskip{5pt}
	\setlength\belowdisplayskip{5pt}
	&\mathbb{E}\left( \left| {{\rm{UI}}_{k,k'}} \right|^2 \right) \notag \\
	 &= \frac{{p_k^p}}{{p_{k'}^p}}N{\left( {\sum\limits_{m \in {{\cal M}_{k'}}} {\sqrt {p_{m,k'}^d{\lambda _{m,k'}}} \frac{{{\beta _{m,k}}}}{{{\beta _{m,k'}}}}} } \right)^2} \!\!\!\!+ \!\!\!\!\sum\limits_{m \in {{\cal M}_{k'}}} {p_{m,k'}^d{\beta _{m,k}}} 	\label{MRC_UIkk_final} \\
	 & = N{\left( {\sum\limits_{m \in {{\cal M}_{k'}}} {\sqrt {p_{m,k'}^d{\lambda _{m,k}}} } } \right)^2} \!\!\!\!+ \!\!\!\! \sum\limits_{m \in {{\cal M}_{k'}}} {p_{m,k'}^d{\beta _{m,k}}} \notag.
\end{align}

The term $\mathbb{E}\left( \left| {{\rm{LS}}_{k}} \right|^2 \right)$ in (\ref{kth_SINR_downlink}) is given by (\ref{MRC_LSk}) at the bottom of next page.
\begin{figure*}[b]
	\hrulefill
	\begin{equation}
		\setlength\abovedisplayskip{5pt}
		\setlength\belowdisplayskip{5pt}
		\label{MRC_LSk}
		\begin{split}
			&\mathbb{E} \left\{ {{{\left| {{\rm{LS}}_{k}} \right|}^2}} \right\}  \\
			&= \mathbb{E}\left\{ {{{\left| {\sum\limits_{m \in {{\cal M}_k}} {{{\left( {{{\bf{g}}_{m,k}}} \right)}^T}{{\left( {{\bf{a}}_{m,k}} \right)}^ * }\sqrt {p_{m,k}^d} }  - {\left| {{\rm{DS}}_{k}^d} \right|}} \right|}^2}} \right\}\\
			&= \mathbb{E} \left\{{\sum\limits_{m \in {{\cal M}_k}} {\frac{{p_{m,k}^d{\lambda _{m,k}}}}{{{\beta _{m,k}}N}}\left( {\sum\limits_{i \in {{\cal Q}_k}} {\sqrt {\frac{{p_i^p}}{{p_k^p}}} {{\left( {{{\bf{g}}_{m,k}}} \right)}^T}} {{\left( {{{\bf{g}}_{m,i}}} \right)}^ * } + \frac{1}{{\sqrt {\tau p_k^p} }}{{\left( {{{\bf{g}}_{m,k}}} \right)}^T}{\bf{N}}_m^p{{\bf{q}}_k}} \right)} }\right\}^2 \\
			&\quad + \mathbb{E}\left\{ \sum\limits_{m \in {{\cal M}_k}} {\sum\limits_{m' \in \left\{ {{{\cal M}_k}\backslash m} \right\}}{\frac{{\sqrt {p_{m,k}^d{\lambda _{m,k}}} }}{{{\beta _{m,k}}\sqrt N }}\frac{{\sqrt {p_{m',k}^d{\lambda _{m',k}}} }}{{{\beta _{m',k}}\sqrt N }}{{\left( {{{\bf{g}}_{m,k}}} \right)}^T}{{\left( {{{\bf{g}}_{m,k}}} \right)}^*}{{\left( {{{\bf{g}}_{m',k}}} \right)}^T}{{\left( {{{\bf{g}}_{m',k}}} \right)}^*}} }\right\} - {\left| {{\rm{DS}}_{k}^d} \right|^2} \\
			& = \sum\limits_{m \in {{\cal M}_k}} {\frac{{p_{m,k}^d{\lambda _{m,k}}}}{{{{\left( {{\beta _{m,k}}} \right)}^2}N}}\left( {N\left( {N + 1} \right){{\left( {{\beta _{m,k}}} \right)}^2} + N\sum\limits_{i \in \left\{ {{{\cal Q}_k}\backslash k} \right\}} {\frac{{p_i^p}}{{p_k^p}}} {\beta _{m,k}}{\beta _{m,i}} + \frac{N}{{\tau p_k^p}}{\beta _{m,k}}} \right)} \\
			& \quad + \sum\limits_{m \in {{\cal M}_k}} {\sum\limits_{m' \in \left\{ {{{\cal M}_k}\backslash m} \right\}} {\frac{{\sqrt {p_{m,k}^d{\lambda _{m,k}}} }}{{{\beta _{m,k}}\sqrt N }}\frac{{\sqrt {p_{m',k}^d{\lambda _{m',k}}} }}{{{\beta _{m',k}}\sqrt N }}{N^2}{\beta _{m,k}}{\beta _{m',k}}} } - {\left( {\sum\limits_{m \in {{\cal M}_k}} {\sqrt {Np_{m,k}^d{\lambda _{m,k}}} } } \right)^2} \\
			& = \sum\limits_{m \in {{\cal M}_k}} {p_{m,k}^d{\lambda _{m,k}}\sum\limits_{i \in {{\cal Q}_k}} {\frac{{p_i^p}}{{p_k^p}}} \frac{{{\beta _{m,i}}}}{{\left( {{\beta _{m,k}}} \right)}}}  + \sum\limits_{m \in {{\cal M}_k}} {\frac{{p_{m,k}^d{\lambda _{m,k}}}}{{\tau p_k^p{\beta _{m,k}}}}} \\
			& = \sum\limits_{m \in {{\cal M}_k}} {p_{m,k}^d{\lambda _{m,k}}\left( {\frac{{\sum\limits_{i \in {{\cal Q}_k}} {p_i^p{\beta _{m,i}}} }}{{p_k^p\left( {{\beta _{m,k}}} \right)}} + \frac{1}{{\tau p_k^p{\beta _{m,k}}}}} \right)}  = \sum\limits_{m \in {{\cal M}_k}} {p_{m,k}^d{\beta _{m,k}}}.
		\end{split}
	\end{equation}
\end{figure*}

Upon substituting (\ref{MRC_DSk}), (\ref{MRC_UIkk_final}), (\ref{MRC_LSk}), and $\mathbb{E} \left\{ {{{\left| {{{\rm N}_k^d}} \right|}^2}} \right\} = 1$ into (\ref{kth_SINR_downlink}), we obtain $\hat \gamma _k^{d}$ in (\ref{MRT_SINR_LB}).
	
\section{Proof of Lemma \ref{MRT_trans}}
\label{MRT_trans_proof}
Upon using the expressions of $\varphi _k$ and ${\theta _{k,k}}$ in (\ref{varphi_k}) and (\ref{theta_k}), we have
\begin{align}
	\setlength\abovedisplayskip{5pt}
	\setlength\belowdisplayskip{5pt}
	&\frac{{{\varphi _k}}}{{{\theta _{k,k}}}} \notag \\
	&= \frac{{\sum\limits_{m \in {{\cal M}_k}} {\sqrt {\tau p_k^pp_{m,k}^d{{\left( {{\beta _{m,k}}} \right)}^2}\!\!\!\!\!\!\prod\limits_{n \in \left\{ {{{\cal M}_k}\backslash m} \right\}} {\left( {\sum\limits_{i \in {{\cal Q}_k}} {\tau p_i^p{\beta _{n,i}}}  + 1} \right)} } } }}{{\prod\limits_{m \in {{\cal M}_k}} {\sqrt {\sum\limits_{i \in {{\cal Q}_k}} {\tau p_i^p{\beta _{m,i}}}  + 1} } }}  \notag \\
	& = \sum\limits_{m \in {{\cal M}_k}} {\sqrt {\frac{{\tau p_k^pp_{m,k}^d{{\left( {{\beta _{m,k}}} \right)}^2}}}{{\sum\limits_{i \in {{\cal Q}_k}} {\tau p_i^p{\beta _{m,i}}}  + 1}}} } \label{first_term}\\
	& = \sum\limits_{m \in {{\cal M}_k}} {\sqrt {p_{m,k}^d{\lambda _{m,k}}} } \notag.
\end{align}

Then, by using (\ref{phi_k}) and (\ref{theta_k}), we obtain
\begin{align}
	\setlength\abovedisplayskip{5pt}
	\setlength\belowdisplayskip{5pt}
	&\frac{{{\phi _{k,k'}}}}{{{\theta _{k,k'}}}} \notag \\
	 & = \frac{{\sum\limits_{m \in {{\cal M}_{k'}}} {\sqrt {\tau p_k^pp_{m,k'}^d{{\left( {{\beta _{m,k}}} \right)}^2}\!\!\!\!\!\!\!\!\prod\limits_{n \in \left\{ {{{\cal M}_{k'}}\backslash m} \right\}}\!\!\! {\left( {\sum\limits_{i \in {{\cal Q}_k}} {\tau p_i^p{\beta _{n,i}}}  + 1} \right)} } } }}{{\!\!\prod\limits_{m \in {{\cal M}_{k'}}}\!\!\!\! {\sqrt {\sum\limits_{i \in {{\cal Q}_k}} {\tau p_i^p{\beta _{m,i}}}  + 1} } }} \notag \\
	& = \sum\limits_{m \in {{\cal M}_{k'}}} {\sqrt {\frac{{\tau p_k^pp_{m,k'}^d{{\left( {{\beta _{m,k}}} \right)}^2}}}{{\sum\limits_{i \in {{\cal Q}_{k}}} {\tau p_i^p{\beta _{m,i}}}  + 1}}} }  \label{second_term}\\ 
	& = \sum\limits_{m \in {{\cal M}_{k'}}} {\sqrt {p_{m,k'}^d{\lambda _{m,k}}} }\notag.
\end{align}

Substituting the expressions in (\ref{first_term}) and (\ref{second_term}) into the SINR's expression in (\ref{MRT_SINR_LB}), we have
\begin{align}
	\setlength\abovedisplayskip{5pt}
	\setlength\belowdisplayskip{5pt}
	&\hat \gamma _k^{{\rm{MRT}}} \notag \\
	& = \frac{{N{{\left( {\frac{{{\varphi _k}}}{{{\theta _{k,k}}}}} \right)}^2}}}{{\sum\limits_{k' = 1}^K {\sum\limits_{m \in {{\cal M}_{k'}}} {p_{m,k'}^d{\beta _{m,k}}} }  + N\sum\limits_{k' \in \left\{ {{{\cal Q}_k}\backslash k} \right\}} {{{\left( {\frac{{{\phi _{k,k'}}}}{{{\theta _{k,k'}}}}} \right)}^2}}  + 1}} 	\label{third_term}.
\end{align}

The term of pilot contamination in (\ref{third_term}) can be written as
\begin{align}
	\setlength\abovedisplayskip{5pt}
	\setlength\belowdisplayskip{5pt}
	& \sum\limits_{k' \in \left\{ {{{\cal Q}_k}\backslash k} \right\}} {{{\left( {\frac{{{\phi _{k,k'}}}}{{{\theta _{k,k'}}}}} \right)}^2}}  \notag \\
	& = \frac{{\sum\limits_{k' \in \left\{ {{{\cal Q}_k}\backslash k} \right\}} {{{\left( {{\phi _{k,k'}}\prod\limits_{j \in \left\{ {{{\cal Q}_k}\backslash \left\{ {k,k'} \right\}} \right\}} {{\theta _{k,j}}} } \right)}^2}} }}{{\prod\limits_{k' \in \left\{ {{{\cal Q}_k}\backslash k} \right\}} {{{\left( {{\theta _{k,k'}}} \right)}^2}} }} 	\label{fourth_term}.
\end{align}

We finally complete this proof by substituting (\ref{fourth_term}) into  (\ref{third_term}).

\section{Proof of Theorem \ref{theorem_MRT}}
\label{Down_proof}
By defining $x_{m,k}^d$ and $x_{i}^p$ as $\ln (p_{m,k}^d{\left( {{\beta _{m,k}}} \right)^2})$ and $\ln (\tau p_i^p)$, we have (\ref{six_term}) at the bottom of this page.
\begin{figure*}[hb]
	\hrulefill
\begin{align}
	\setlength\abovedisplayskip{5pt}
	\setlength\belowdisplayskip{5pt}
	  {\varphi _k}\prod\limits_{k' \in \left\{ {{{\cal Q}_k}\backslash k} \right\}} {\left( {{\theta _{k,k'}}} \right)}
	= {\sum\limits_{m \in {{\cal M}_k}} {\sqrt {{e^{x_{m,k}^d}}{e^{x_k^p}}\prod\limits_{n \in \left\{ {{{\cal M}_k}\backslash m} \right\}} {\left( {\sum\limits_{i \in {{\cal Q}_k}} {{e^{x_i^p}}{\beta _{n,i}}}  + 1} \right)} } } } \prod\limits_{k' \in \left\{ {{{\cal Q}_k}\backslash k} \right\}} {\left( {\prod\limits_{m \in {{\cal M}_{k'}}} {\sqrt {\sum\limits_{i \in {{\cal Q}_k}} {e^{x_i^p}{\beta _{m,i}}}  + 1} } } \right)}.  \label{six_term}
\end{align}
\end{figure*}

Then, by taking the logarithm of both sides in (\ref{six_term}), we obtain
\begin{align}
	\setlength\abovedisplayskip{5pt}
	\setlength\belowdisplayskip{5pt}
 &\ln \left( {{\varphi _k}\prod\limits_{k' \in \left\{ {{{\cal Q}_k}\backslash k} \right\}} {\left( {{\theta _{k,k'}}} \right)} } \right)  \notag \\
 &=\ln \left( {\sum\limits_{m \in {{\cal M}_k}} {\sqrt {{e^{x_{m,k}^d}}{e^{x_k^p}}\prod\limits_{n \in \left\{ {{{\cal M}_k}\backslash m} \right\}} {\left( {\sum\limits_{i \in {{\cal Q}_k}} {{e^{x_i^p}}{\beta _{n,i}}}  + 1} \right)} } } } \right) \notag \\
 &  \quad +  \sum\limits_{k' \in \left\{ {{{\cal Q}_k}\backslash k} \right\}} {\sum\limits_{m \in {{\cal M}_{k'}}} {\ln \sqrt {\left( {\sum\limits_{i \in {{\cal Q}_k}} {{e^{x_i^p}}{\beta _{m,i}}}  + 1} \right)} } }\label{seven} \\
 & \buildrel \Delta \over = F({\bf{x}}) \notag,
\end{align}
where $\bf x$ is a vector that collects $x_i^p$ and $x_{m,k}^d$, $\forall i \in {\cal{Q}}_k$ and $\forall m \in {\cal{M}}_k$.

In the following, we prove $F({\bf{x}})$ is a convex function of $\bf x$. Firstly, as ${ \sqrt {\left( {\sum\limits_{i \in {{\cal Q}_k}} {{e^{x_i^p}}{\beta _{m,i}}}  + 1} \right)} }$ is a log-convex function of $x_i^p$, $\forall i \in {\cal{Q}}_k $, we can prove that $\ln { \sqrt {\left( {\sum\limits_{i \in {{\cal Q}_k}} {{e^{x_i^p}}{\beta _{m,i}}}  + 1} \right)} }$ is a convex function. Then, it is readily shown that the second term in $F(\bf{x})$ is a convex function by using the property that the sum of convex functions is also convex. Furthermore, it is readily to prove that $\left( {{\sqrt {{e^{x_{m,k}^d}}{e^{x_k^p}}\prod\limits_{n \in \left\{ {{{\cal M}_k}\backslash m} \right\}} {\left( {\sum\limits_{i \in {{\cal Q}_k}} {{e^{x_i^p}}{\beta _{n,i}}}  + 1} \right)} } } } \right)$ in the first term is a log-convex function. Finally, we can prove that $\left( {\sum\limits_{m \in {{\cal M}_k}} {\sqrt {{e^{x_{m,k}^d}}{e^{x_k^p}}\prod\limits_{n \in \left\{ {{{\cal M}_k}\backslash m} \right\}} {\left( {\sum\limits_{i \in {{\cal Q}_k}} {{e^{x_i^p}}{\beta _{n,i}}}  + 1} \right)} } } } \right)$ is a log-convex function, since the sum of log-convex functions is also a log-convex function. As a result, we complete the proof that $F({\bf{x}})$ is a convex function of $\bf x$ by using the properties of convex functions.

Upon using the Jensen's inequality, we have
\begin{equation}
	\setlength\abovedisplayskip{5pt}
	\setlength\belowdisplayskip{5pt}
	\label{eight}
	F({\bf{x}}) \ge \ln({c_k}) + \sum\limits_{m \in {{\cal M}_k}} {{{{a_{m,k}}}x_{m,k}^d}} + \sum\limits_{i \in {{\cal Q}_k}} {{{{b_i}} x_i^p}},
\end{equation}
where $c_k$, $a_{m,k}$, and $b_i$ are given in (\ref{MRT_theorem_ek}), (\ref{MRT_theorem_amk}), and (\ref{MRT_theorem_bi}), respectively.
Finally, we complete this proof by taking the exponential operation on both sides of (\ref{eight}).
\end{appendices}

\bibliographystyle{IEEEtran}
\bibliography{myref}

\end{document}